\documentclass[aps, prd, twocolumn, notitlepage, nofootinbib]{revtex4-1}
\usepackage{graphicx}
\usepackage{float}
\usepackage{amsmath}
\usepackage{color}
\usepackage{tensor}
\usepackage{epstopdf}
\usepackage{xcolor}
\usepackage{wasysym}
\usepackage{hyperref}
\hypersetup{
    colorlinks,
    linkcolor={blue!50!black},
    citecolor={red!50!black},
    urlcolor={blue!80!black}
}

\usepackage{multibib}

\newcommand{\be}{\begin{equation}}
\newcommand{\ee}{\end{equation}}
\newcommand{\ba}{\begin{eqnarray}}
\newcommand{\ea}{\end{eqnarray}}

\newcommand{\beq}{\begin{equation}}
\newcommand{\eeq}{\end{equation}}
\newcommand{\beqa}{\begin{eqnarray}}
\newcommand{\eeqa}{\end{eqnarray}}
\newcommand{\nn}{\nonumber}
\graphicspath{{plots/}}

\begin{document}

\title{Thermodynamics of $z=4$ Ho\v{r}ava-Lifshitz black holes}

\author{Mohammad Bagher Jahani Poshteh}
\email{jahani@ipm.ir}
\affiliation{School of Physics, Institute for Research in Fundamental Sciences (IPM), P.O. Box 19395-5531, Tehran, Iran}

\author{Robert B. Mann}
\email{rbmann@uwaterloo.ca}
\affiliation{Department of Physics and Astronomy, University of Waterloo, 200 University Ave W, Waterloo, Ontario, N2L 3G1 Canada}
\affiliation{Perimeter Institute for Theoretical Physics, 31 Caroline St. N., Waterloo, Ontario, N2L 2Y5 Canada}

%

\pacs{04.50.Gh, 04.70.-s, 05.70.Ce}

\begin{abstract}
Thermodynamics of $z=4$ Ho\v{r}ava-Lifshitz black holes in 3+1 dimensions is studied in extended phase space. By using the scaling argument we find the Smarr relation and the first law for the black hole solutions of $z=4$ Ho\v{r}ava-Lifshitz gravity. We find that it is necessary to take into account the variation of dimensionful parameters of the theory in the first law. We find that the reverse isoperimetric inequality can be violated for spherical, flat, and hyperbolic horizons and in all such cases we have black holes for which the specific heat at constant pressure and volume are positive. This provides a counterexample to a recent conjecture stating that black holes violating the reverse isoperimetric inequality are thermodynamically unstable. We find for $z=4$ Ho\v{r}ava-Lifshitz black holes with hyperbolic horizons that there are two critical points: one showing Van der Waals behavior, the other reverse Van der Waals behavior.
\end{abstract}

\maketitle

\section{Introduction}

The laws of black hole mechanics~\cite{bardeen1973} have had a profound impact on our understanding of black hole physics. They provided the physics community with robust clues as to the thermal properties of black holes. These laws, supplemented by the proposals for black hole entropy~\cite{bekenstein1973} and temperature~\cite{hawking1974}, motivated the study of black hole thermodynamics for the past half-century~\cite{Unruh:1980cg,deNova:2018rld,Davies:1989ey,Chamblin:1999tk,Chamblin:1999hg,Cai:1998ep,Aman:2003ug,Shen:2005nu,Poshteh:2013pba,Poshteh:2015ova,Niu:2011tb,Banerjee:2011cz,Banerjee:2010bx,Hennigar:2015cja,Appels:2019vow}.

Of particular interest is the thermodynamics of black holes in asymptotically AdS spacetime, in which a negative cosmological constant is present. This is due, in part, to the role these black holes play in the AdS/CFT correspondence~\cite{maldacena1999}, and their rich thermodynamic phase structure~\cite{hawking1983,Kubiznak:2016qmn}. In asymptotically AdS spacetime, an extension of the laws of black hole mechanics~\cite{bardeen1973} implies that the first law should be modified to include the variation of the cosmological constant~\cite{kastor2009}, consistent with Eulerian scaling \cite{Kubiznak:2016qmn}. This way, one regards the negative cosmological constant as the pressure and its conjugate quantity as thermodynamic volume in an extended phase space~\cite{dolan2011}.

The investigations of black hole thermodynamics in extended phase space have revealed a deep resemblance between charged AdS black holes and Van der Waals fluid~\cite{kubizvnak2012}. Furthermore, studies of higher dimensional spinning black holes in AdS background have shown the phenomena of reentrant phase transitions~\cite{altamirano2013} and the existence of a triple point~\cite{altamirano2014}. These features have their counterparts in thermodynamics of multicomponent liquid mixtures~\cite{narayanan1994}, polymers \cite{Dolan:2014vba} , and superfluids \cite{Hennigar:2016xwd}, and have the potential to pave our way to understanding of the microstructure of black holes
\cite{Wei:2019uqg}.

In this paper we study the thermodynamics of superrenormalizable Ho\v{r}ava-Lifshitz black holes. As a candidate of quantum gravity theory, Ho\v{r}ava-Lifshitz gravity has attracted considerable attention since its original proposal~\cite{hovrava2009:1,hovrava2009:2}. By relaxing Lorentz invariance in the UV limit, Ho\v{r}ava's theory is power-counting (super)renormalizable and reduces to the relativistic theory of Einstein in the IR limit.

In Ho\v{r}ava's scenario, space and time obey the Lifshitz-type anisotropic scaling
\be
\textbf{x}\rightarrow b\textbf{x}, \qquad t\rightarrow b^z t,
\ee
with dynamical critical exponent $z$. In $D+1$ dimensional spacetime the theory becomes power-counting renormalizable upon taking $z=D$. Black hole solutions of $z=D=3$ Ho\v{r}ava-Lifshitz gravity have been obtained~\cite{lu2009,cai2009:1,kehagias2009,ghodsi2010} and their thermodynamics investigated~\cite{myung2010,majhi2010,Majhi:2012fz,poshteh2017,ma2017,pourhassan2019}.

However much less is known about the
the $z=4$ case~\cite{cai2009:2,chen2009,liu2013}, notwithstanding its greater importance. Numerical simulations in lattice quantum gravity, in the framework of causal dynamical triangulations, suggest that the spectral dimension of spacetime is $d_s=1.80\pm0.25$ in the short-distance limit~\cite{ambjorn2005}. In spacetimes with anisotropic scaling, the spectral dimension is given by $d_s=1+D/z$~\cite{hovrava2009:3} -- for $D=3$, these results favor $z=4$ more than $z=3$. Further arguments supporting $z=4$ have also been made~\cite{hovrava2009:2,cai2009:2}.

Our study of $z=4$ Ho\v{r}ava-Lifshitz gravity shows that in the superrenormalizable theory, there are two critical points of black hole thermodynamics. This is unlike the $z=3$ case where no criticality exists~\cite{mo2015}. The critical behavior of $z=4$ Ho\v{r}ava-Lifshitz black holes is peculiar, exhibiting some interesting features. One example is that of critical points with Van der Waals behavior characterized by an inverted swallowtail structure. At higher temperatures and pressures there exists another critical point with reverse Van der Waals behavior~\cite{frassino2014,Kubiznak:2016qmn}.

We also study the reverse isoperimetric inequality \cite{cvetivc2011} for three cases of spherical $k=1$, flat $k=0$, and hyperbolic $k=-1$ horizons. In all three cases this inequality can be violated for $z=4$ Ho\v{r}ava-Lifshitz black holes. Furthermore, in such situations, the specific heat at constant pressure $C_P$ and constant volume $C_V$ can be both positive for black holes. We therefore obtain a counterexample to recent conjectures stating that black holes violating the reverse isoperimetric inequality are thermodynamically unstable~\cite{johnson2020,cong2019}.

The outline of our paper is as follows. In the next section we review $z=4$ Ho\v{r}ava-Lifshitz gravity and present its black hole solutions. The Smarr relation and the first law of these black holes are obtained in section~\ref{sec:smarr}, by using the scaling relation. In section~\ref{sec:rii} we study the reverse isoperimetric inequality and show that $z=4$ Ho\v{r}ava-Lifshitz black hole can violate the inequality while they are thermodynamically stable. In section~\ref{sec:crit} we investigate the critical behavior of the $z=4$ Ho\v{r}ava-Lifshitz black hole and show that they exhibit both Van der Waals and reverse Van der Waals behavior. We conclude our paper in section~\ref{sec:con}. The basic thermodynamic quantities and the first law of $k=0$ $z=4$ Ho\v{r}ava-Lifshitz black hole are provided in an appendix.

\section{$z=4$ Ho\v{r}ava-Lifshitz gravity}\label{sec:z4hl}

In the ADM decomposition the metric is given by
\begin{equation}
	ds^{2}=-N^{2}c^{2}dt^{2}+g_{ij}(dx^{i}-N^{i}dt)(dx^{j}-N^{j}dt),
\end{equation}
with $i,j=1,2,3.$ Here, $c$, $N$, and $N^{i}$ are the speed of light, the lapse, and the shift functions, respectively. Given the degree of anisotropy $z=4$, we have, in units of spatial length,
\begin{equation}
	[dx]=[dt]^{\frac{1}{4}}=[c]^{-\frac{1}{3}}=[N_{i}]^{-\frac{1}{3}}=1.
\end{equation}
The remaining fields, $g_{ij}$ and $N$, are dimensionless.

The kinetic term of the Lagrangian is~\cite{hovrava2009:2}
\begin{equation}
	\mathcal{L}_{K}=\frac{2}{\kappa^{2}}\sqrt{g}N(K_{ij}K^{ij}-\lambda K^{2}),
	\label{kinlag}
\end{equation}
where $K_{ij}=\frac{1}{2N}(\dot{g}_{ij}-\nabla_{i}N_{j}-\nabla_{j}N_{i})$, and $\kappa$ and $\lambda$ are coupling constants. From now on, we take $\lambda=1$ so that the theory returns to general relativity in the low energy limit. Note also that $\kappa$ is of dimension -1/2 in units of spatial length. The potential Lagrangian is~\cite{hovrava2009:2}
\begin{equation}
	\mathcal{L}_{V}=\frac{\kappa^{2}}{8}\sqrt{g}N E^{ij}\mathcal{G}_{ijkl}E^{kl},
	\label{potlag}
\end{equation}
where $\mathcal{G}_{ijkl}=\frac{1}{2}(g_{ik}g_{jl}+g_{il}g_{jk}-g_{ij}g_{kl})$ is the inverse De Witt metric. In order to obtain $E^{ij}$, we have 
\begin{equation}
	E^{ij}=\frac{1}{\sqrt{g}}\frac{\delta W[g_{ij}]}{\delta g_{ij}}, \label{eij}
\end{equation}
from the detailed balanced condition, where $W$ is the action of a 3-dimensional relativistic theory with Euclidean signature. We take $W$ to be the Euclidean sector of the action proposed by Bergshoeff, Hohm, and Townsend (BHT)~\cite{bergshoeff2009}
\begin{equation}
	W=\mu\int d^{3}x\sqrt{g}\left[R-2\Lambda_{W}+\frac{1}{m^{2}}\left(R_{ij}R^{ij}-\frac{3}{8}R^{2}\right)\right], \label{wnmg}
\end{equation}
where $\Lambda_{W}, R_{ij}$, and $R$ are respectively the cosmological constant, Ricci tensor, and Ricci scalar of the 3-dimensional special theory (i.e.\ BHT). We have  $[R]=[\Lambda_{W}]= [{\rm length}]^{-2}$; $\mu$ and $m$ are coupling coefficients both of dimension -1 in units of spatial length.

In order to recover general relativity in the long-distance limit, the speed of light $c$ and the Newton coupling constant $G_N$ should be related to the effective coupling constants of Ho\v{r}ava-Lifshitz theory through
\be
c=\frac{\mu\kappa^2}{4}\sqrt{\frac{\Lambda_{W}}{-2}}, \qquad G_N=\frac{\kappa^2c}{32\pi}.
\ee
This clearly shows that the cosmological constant $\Lambda_{W}$ has to be negative.

By using eqs. (\ref{eij}) and (\ref{wnmg}) we write
\begin{equation}
	E^{ij}=-\mu\left(G^{ij}+\Lambda_{W}g^{ij}+\frac{1}{m^{2}}L^{ij}\right),
\end{equation}
where $G^{ij}=R^{ij}-\frac{1}{2}g^{ij}R$ is the Einstein tensor and
\begin{eqnarray}
	L^{ij}&=&\frac{1}{4}\left(g^{ij}\nabla^{2}-\nabla^{i}\nabla^{j}\right)R-\frac{3}{4}R\left(R^{ij}-\frac{1}{4}g^{ij}R\right) \nn\\
	&+&\nabla^{2}G^{ij}+2\left(R^{imjn}-\frac{1}{4}g^{ij}R^{mn}\right)R_{mn}.
\end{eqnarray}
The total Lagrangian, using (\ref{kinlag}) and (\ref{potlag}), is
\begin{eqnarray}
	\mathcal{L}&=&\mathcal{L}_{K}+\mathcal{L}_{V}=\frac{\sqrt{g}N\kappa^{2}\mu^{2}}{16}\biggl\{\frac{32}{\kappa^{4}\mu^{2}}\left(K^{ij}K_{ij}-K^{2}\right)+\frac{R^{2}}{4} \nn \\
	&+&\Lambda_{W}\left(3\Lambda_{W}-R\right)-2G^{ij}G_{ij}-\frac{1}{m^{2}}\left[L\left(R-2\Lambda_{W}\right)\right. \nn \\
	&+&\left.4G^{ij}L_{ij}\right]+\frac{1}{m^{4}}\left(L^{2}-2L^{ij}L_{ij}\right)\biggr\},
	\label{totlag}
\end{eqnarray}
in which
\begin{equation}
	L=g^{ij}L_{ij}=\frac{1}{2}R^{ij}R_{ij}-\frac{3}{16}R^{2}\; .
\end{equation}

We want to obtain spherically symmetric black hole solutions of the form
\begin{equation}
	ds^{2}=-g^2(r) f(r)c^{2}dt^{2}+\frac{dr^{2}}{f(r)}+r^{2}d\Omega_{k}^{2},
	\label{ssm}
\end{equation}
where $d\Omega_{k}^{2}$ is the line element of 2-dimensional Einstein space with scalar curvature $2k$. Putting the metric (\ref{ssm}) into eq. (\ref{totlag}) 
and performing the variational principle, we find that $g(r)=1$ and\footnote{In fact, there are two solutions
	\begin{equation}
		f(r)=k+2m^{2}r^{2}\left(1\pm\sqrt{1+\frac{\Lambda_{W}}{m^{2}}\pm\frac{\sqrt{c_{+}r}}{\Lambda_{W}\mu m^{2}r^{2}}}\right), \nn
	\end{equation}
	but only the minus branch has a well-defined $m^2\rightarrow\infty$ limit. In this limit the Lagrangian and the metric recover the $z=3$ theory ~\cite{cai2009:2}.
}~\cite{cai2009:2}
\begin{equation}
	f(r)=k+2m^{2}r^{2}\left(1-\sqrt{1+\frac{\Lambda_{W}}{m^{2}}-\frac{\sqrt{c_{+}r}}{\Lambda_{W}\mu m^{2}r^{2}}}\right), \label{metricfunc}
\end{equation}
where
\begin{equation}
	c_{+}=\frac{\mu^{2}\Lambda_{W}^{2}}{r_{+}}\left(\frac{k^{2}}{4m^{2}r_{+}^{2}}+k-\Lambda_{W}r_{+}^{2}\right)^{2},
\end{equation}
is an integration constant defined in terms of $r_+$, which is the outermost solution of $f(r)=0$, corresponding to the radius of the event horizon. Inspection of (\ref{metricfunc}) indicates that the negative cosmological constant and the coupling coefficient $m$ must satisfy
\be
\frac{|\Lambda_{W}|}{m^{2}}\leq 1, \label{eqn:lambdacon}
\ee
so that the metric function is well behaved at large distances.

The mass, temperature, and the entropy of the black hole solution have been previously obtained\footnote{We respect the notation of~\cite{bergshoeff2009}; the coupling $m$ is the so called ``relative'' mass parameter of BHT massive gravity and is related to the parameter $\tilde{\beta}$ of~\cite{cai2009:2} through $m^2=\mu\Lambda_{W}^{2}/(4\tilde{\beta})$.}~\cite{cai2009:2} 
\begin{eqnarray}
	M&=&\frac{\mu^{2}\kappa^{2}\Omega_{k}}{256m^{4}r_+^{5}}\left(k^{2}+4km^{2}r_+^{2}-4\Lambda_{W}m^{2}r_+^{4}\right)^2, \label{mass} \\
	T&=&-\frac{5k^{2}+4km^{2}r_+^{2}+12\Lambda_{W}m^{2}r_+^{4}}{16k\pi r_++32\pi m^{2}r_+^{3}}, \label{temp} \\
	S&=&S_0-\frac{\pi\mu^{2}\kappa^{2}\Omega_{k}}{64m^{4}r_+^{4}}\biggl[k^{3}+12k^{2}m^{2}r_+^{2} +16\Lambda_{W}m^{4}r_+^{6}\nn\\
	&+&16km^{2}r_+^{4}\ln\left(\sqrt{-\Lambda_{W}}r_+\right)(\Lambda_{W}-2m^{2})\biggr], \label{ent}
\end{eqnarray}
where $\Omega_{k}$ denotes the volume of the 2-dimensional Einstein space. We shall set the integration constant $S_0=0$ (\ref{ent}) without loss of generality.
Note that the entropy is not proportional to the area of the horizon.

\section{Smarr formula and the first law} \label{sec:smarr}

Let us redefine the mass (\ref{mass}) and entropy (\ref{ent}) as
\be
M\rightarrow \frac{64M}{\mu^{3}\kappa^{4}\Omega_{k}}, \qquad S\rightarrow \frac{64S}{\mu^{3}\kappa^{4}\Omega_{k}}, \label{eqn:transformations}
\ee
so that the mass has dimension of length and the entropy has dimension of $[{\rm length}]^2$, where $[\kappa] = -1/2$. 

Furthermore, recall that $\Lambda_{W}$ is the cosmological constant of the 3-dimensional special theory with Euclidean signature and is related to the effective cosmological constant by $\Lambda=3\Lambda_{W}/2$~\cite{hovrava2009:2}. In extended phase space we define the pressure as
\be
P=-\frac{\Lambda}{8\pi}=-\frac{3\Lambda_W}{16\pi}. \label{eqn:pressure}
\ee
Using this definition, we can rewrite eqs. (\ref{mass}), (\ref{temp}), and (\ref{ent}) as the following
\begin{eqnarray}
	M&=&\frac{\left(3 k^2+12 k m^2 r_+^2+64 \pi  m^2 P r_+^4\right)^2}{36 \mu\kappa^{2} m^4 r_+^5}, \label{eqn:mass} \\
	T&=&-\frac{5 k^2+4 k m^2 r_+^2-64 \pi  m^2 P r_+^4}{16 \pi  k r_++32 \pi  m^2 r_+^3}, \label{eqn:temp} \\
	S&=&-\frac{\pi}{3 \mu\kappa^{2} m^4r_+^4}\biggl[3 k^3+36 k^2 m^2 r_+^2-32 k m^2 r_+^4 \left(3 m^2\right. \nn\\
	&+&\left.8 \pi  P\right) \ln \left(4 \sqrt{\frac{\pi P}{3}} r_+\right)-256 \pi  m^4 P r_+^6\biggr]. \label{eqn:ent}
\end{eqnarray}

Before we proceed to find the Smarr relation and the first law, we note that the condition \eqref{eqn:lambdacon} can be translated to a condition on the pressure. By using the relation \eqref{eqn:pressure} we find an upper limit 
\be
P \leq P_{max} \equiv \frac{3m^2}{16 \pi}, \label{eqn:press:ul}
\ee
on the pressure, similar to what happens for Lovelock gravity~\cite{frassino2014}.

The Smarr relation for Lifshitz black holes can be non-trivial to demonstrate \cite{Brenna:2015pqa}.
We can obtain it for $z=4$ Ho\v{r}ava-Lifshitz black holes via a consideration of the scaling properties of their thermodynamic parameters.
Consider a function $f(x,y)$ that obeys the scaling relation $f(\alpha^{p}x,\alpha^{q}y)=\alpha^{r}f(x,y)$. Euler's theorem on homogeneous functions then states that~\cite{kastor2009}
\begin{equation}
	rf(x,y)=p\left.\frac{\partial f}{\partial x}\right|_{y}x+q\left.\frac{\partial f}{\partial y}\right|_{x}y.
\end{equation}
If we reverse eq. (\ref{eqn:ent}) to find $r_+$, and substitute it in eq. (\ref{eqn:mass}), we find the mass as a function of $S$, $\mu$, $\kappa$, $m$, and $P$. These quantities would scale according to their respective dimensions: $M\propto\alpha$, $S\propto\alpha^2$, $m \propto \mu \propto \alpha^{-1}$, $\kappa \propto \alpha^{-1/2}$ and $P\propto\alpha^{-2}$. Using Euler's theorem we obtain
\begin{equation}
	M=2ST - 2PV - mH - \mu U - \frac{1}{2}\kappa Y, \label{eqn:smarr}
\end{equation}
where
\begin{eqnarray}
	T&=&\left.\frac{\partial M}{\partial S}\right|_{\mu,\kappa,m,P}, \quad V=\left.\frac{\partial M}{\partial P}\right|_{\mu,\kappa,S,m}, \nn\\
	H&=&\left.\frac{\partial M}{\partial m}\right|_{\mu,\kappa,S,P}, \quad U=\left.\frac{\partial M}{\partial \mu}\right|_{\kappa,S,m,P}, \nn\\
	Y&=&\left.\frac{\partial M}{\partial \kappa}\right|_{\mu,S,m,P}.
\end{eqnarray}
with $T$ the temperature, $V$ the thermodynamic volume, and $H$, $U$, and $Y$ are the respective thermodynamic conjugates of $m$, $\mu$, and $\kappa$.

Although it is possible to solve \eqref{eqn:temp} for $r_+$ in terms of the other thermodynamic variables, the resultant expressions are cumbersome. It is easier to use\footnote{ The calculations considerably simplify if $k=0$, as we show in appendix \ref{app:k0case}.}
\begin{eqnarray}
	T&=&\left.\frac{\partial M}{\partial r_+}\right|_{\mu,\kappa,m,P}\left.\frac{\partial r_+}{\partial S}\right|_{\mu,\kappa,m,P}, \label{tempinf}\\
	V&=&\left.\frac{\partial M}{\partial P}\right|_{\mu,\kappa,r_+,m}+\left.\frac{\partial M}{\partial r_+}\right|_{\mu,\kappa,m,P}\left.\frac{\partial r_+}{\partial P}\right|_{\mu,\kappa,S,m},\label{volumeinf}\\
	H&=&\left.\frac{\partial M}{\partial m}\right|_{\mu,\kappa,r_+,P}+\left.\frac{\partial M}{\partial r_+}\right|_{\mu,\kappa,m,P}\left.\frac{\partial r_+}{\partial m}\right|_{\mu,\kappa,S,P}, \\
	U&=&\left.\frac{\partial M}{\partial \mu}\right|_{\kappa,r_+,m,P}+\left.\frac{\partial M}{\partial r_+}\right|_{\mu,\kappa,m,P}\left.\frac{\partial r_+}{\partial \mu}\right|_{\kappa,S,m,P}, \\
	Y&=&\left.\frac{\partial M}{\partial \kappa}\right|_{\mu,r_+,m,P}+\left.\frac{\partial M}{\partial r_+}\right|_{\mu,\kappa,m,P}\left.\frac{\partial r_+}{\partial \kappa}\right|_{\mu,S,m,P}.
\end{eqnarray}
Note how the redefinitions \eqref{eqn:transformations} affect $V$, $H$, $U$, and $Y$, but not $T$.
One can easily check that eq. (\ref{tempinf}) gives the temperature (\ref{eqn:temp}) which is also obtained in \cite{chen2009} by investigating the fermion tunneling effects \cite{Kerner:2007rr}. We can find the thermodynamic volume and the quantities conjugate to $m$, $\mu$, and $\kappa$ respectively as
\begin{widetext}
	\begin{eqnarray}
		V&=&\frac{32 \pi \left(3 k^2+12 k m^2 r_+^2+64 \pi m^2 P r_+^4\right)}{9\mu \kappa^2 m^2 r_+}+\frac{\left(5 k^2+4 k m^2 r_+^2-64 \pi  m^2 P r_+^4\right)}{3 \mu \kappa^2 m^2 P r_+ \left(k+2 m^2 r_+^2\right)} \nn\\
		&\qquad&\qquad\qquad\qquad\qquad\qquad\times\left[3 k m^2 +8 \pi  k P + 8 \pi  k P\ln \left(\frac{16}{3} \pi  P r_+^2\right) +16 \pi m^2 P r_+^2\right],\label{eqn:volume}\\
		H&=&\frac{-k^2\left(3 k^2+12 k m^2 r_+^2+64 \pi m^2 P r_+^4\right)}{3 \mu \kappa^2 m^5 r_+^5}+\frac{k\left(5k^2+4 k m^2 r_+^2-64 \pi  m^2 P r_+^4\right)}{12 \mu \kappa^2 m^5 r_+^5 \left(k+2 m^2 r_+^2\right)}\nn\\
		&\qquad&\qquad\qquad\qquad\qquad\qquad\times \left[3k^2+18k m^2 r_+^2-64 \pi m^2 P r_+^4 \ln \left(\frac{16}{3} \pi  P r_+^2\right)\right], \label{eqn:h}\\
		U&=&\frac{- \left(3 k^2+12 k m^2 r_+^2+64 \pi m^2 P r_+^4\right)^2}{36\mu^2 \kappa^2 m^4 r_+^5}+\frac{\left(5 k^2+4 k m^2 r_+^2-64 \pi  m^2 P r_+^4\right)}{48 \mu^2 \kappa^2 m^4 r_+^5 \left(k+2 m^2 r_+^2\right)} \nn\\
		&\qquad&\qquad\qquad\qquad\qquad\qquad\times\left[3 k^3+36 k^2 m^2 r_+^2-32 k m^2 r_+^4 \left(3 m^2+8 \pi  P\right) \ln \left(4 \sqrt{\frac{\pi P}{3}} r_+\right)-256 \pi  m^4 P r_+^6\right],\label{eqn:u}\\
		Y&=&\frac{- \left(3 k^2+12 k m^2 r_+^2+64 \pi m^2 P r_+^4\right)^2}{18\mu \kappa^3 m^4 r_+^5}+\frac{\left(5 k^2+4 k m^2 r_+^2-64 \pi  m^2 P r_+^4\right)}{24 \mu \kappa^3 m^4 r_+^5 \left(k+2 m^2 r_+^2\right)} \nn\\
		&\qquad&\qquad\qquad\qquad\qquad\qquad\times\left[3 k^3+36 k^2 m^2 r_+^2-32 k m^2 r_+^4 \left(3 m^2+8 \pi  P\right) \ln \left(4 \sqrt{\frac{\pi P}{3}} r_+\right)-256 \pi  m^4 P r_+^6\right],\label{eqn:y}
	\end{eqnarray}
\end{widetext}
where we note that $[V]=[{\rm length}]^3$, $[H]=[U]=[{\rm length}]^2$, and $[Y^2]=[{\rm length}]^3$.

Using \eqref{eqn:volume} -- \eqref{eqn:y}, along with the entropy \eqref{eqn:ent} and the temperature \eqref{eqn:temp}, we find that the Smarr relation \eqref{eqn:smarr} for $z=4$ Ho\v{r}ava-Lifshitz black holes is indeed satisfied. We note that one should take into account the variation of all the dimensionful parameters, so that one obtains the exact formula for the mass. It is also straightforward to check that the first law
\begin{equation}
	dM=TdS+VdP+Hdm+Ud\mu+Yd\kappa ,
\end{equation}
holds.

\begin{figure}[htp]
	\centering
	\includegraphics[width=0.45\textwidth]{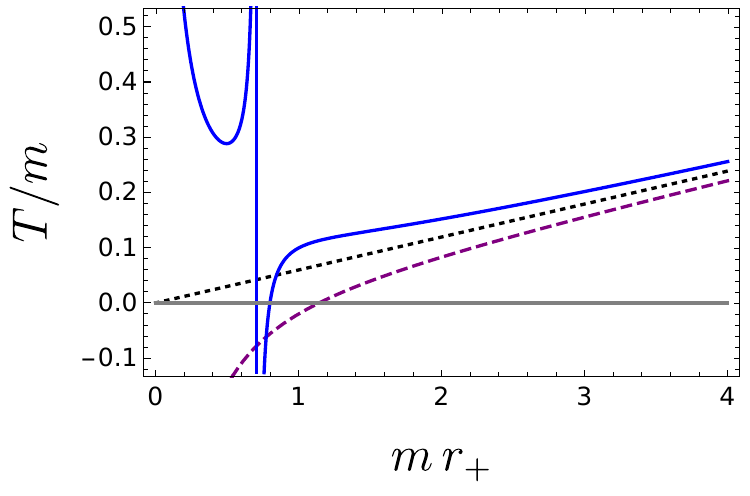}
	\caption{The temperature as a function of horizon radius, with $P_c<P= 0.0298416 m^2<\tilde{P}_c$. The dotted (black) line corresponds to $k=0$, the dashed (purple) curve to $k=1$, which is zero for $m r_+ = 1.14244$. The solid (blue) curve corresponds to $k=-1$; it vanishes for $m r_+ = 0.799057$ and diverges at $m r_+ = 1/\sqrt{2}$ (vertical blue line).}
	\label{fig:temp}
\end{figure}

In the rest of this section we study the temperature of $z=4$ Ho\v{r}ava-Lifshitz black holes in more detail. The first point to note is that
the temperature \eqref{eqn:temp} vanishes at 
\begin{equation}\label{eqn:rmin}
	r_{+,\textrm{min}}= \sqrt{\frac{km + \sqrt{80\pi P + m^2}}{32\pi P m}} ,
\end{equation}
for $k=\pm 1$. For $k=1$, this indicates a minimal size for spherical black holes, illustrated by the intersection of the dashed line with the axis in figure \ref{fig:temp}. 
For the planar horizon case, $k=0$, the temperature is
\be
T_{k=0}=2P r_+,
\ee
which is always positive and increasing with the horizon radius.

Hyperbolic horizons ($k=-1$) are much more interesting. In this case, from eq. (\ref{eqn:rmin}) we see that there is a branch of large (cold) black holes of minimal radius. However there is also a branch of small (hot) black holes with $m r_+ < 1/\sqrt{2}$ (in the following we take $m$ to be positive), with the temperature diverging as $m r_+ \to 1/\sqrt{2}$. Solving the relations
\be\label{eqn:inflection:tem}
\frac{\partial T_{k=-1}}{\partial r_+}=0, \qquad \frac{\partial^2 T_{k=-1}}{\partial r_+^2}=0,
\ee
we find two inflection points at
\ba
r_{+,c}&=&\sqrt{\frac{1}{2m^2} \left(6+\sqrt{21}\right)}, \nn\\ \tilde{r}_{+,c}&=&\sqrt{\frac{1}{2m^2} \left(6-\sqrt{21}\right)}, \label{eqn:crit:rad}
\ea
provided the pressures respectively satisfy 
\ba
P_c&=&\frac{\left(69-14 \sqrt{21}\right) m^2}{720 \pi }, \nn\\ \tilde{P}_c&=&\frac{\left(69+14 \sqrt{21}\right) m^2}{720 \pi }, \label{eqn:crit:pres}
\ea
at these points. Note that $P_c<\tilde{P}_c<P_{max}$. This feature of $z=4$ Ho\v{r}ava-Lifshitz black holes does not occur for $z=3$ \cite{cao2011,poshteh2017}. The corresponding expressions for the temperature are 
\ba
T_c&=&\frac{\sqrt{\left(33-7 \sqrt{21}\right)m^2}}{12 \pi }, \nn\\ \tilde{T}_c&=&\frac{\sqrt{\left(33+7 \sqrt{21}\right)m^2}}{12 \pi }, \label{eqn:crit:tem}
\ea
upon inserting the values of the radius and pressure from eqs. \eqref{eqn:crit:rad} and \eqref{eqn:crit:pres} into eq. \eqref{eqn:temp}.

In figure \ref{fig:temp} we plot the temperature (\ref{eqn:temp}) as a function of the event horizon radius for the $k=1,\, 0,\, -1$ geometries, with $P_c<P<\tilde{P}_c$. For the $k=1,\, -1$ cases the temperature can be negative if $r_+ < r_{+,\textrm{min}}$ (and $1/\sqrt{2} < m r_+$ for $k=-1$); we regard such regions as unphysical. We see that for the hyperbolic horizon, $k=-1$, the temperature possesses a single minimum in the small black hole branch to the left of the divergent point. The behaviour of the temperature for $k=1,\, -1$ cases is qualitatively similar to the corresponding cases for charged topological $z=3$ Ho\v{r}ava-Lifshitz black holes~\cite{cao2011}.

\begin{figure*}[htp]
	\centering
	\includegraphics[width=0.496\textwidth]{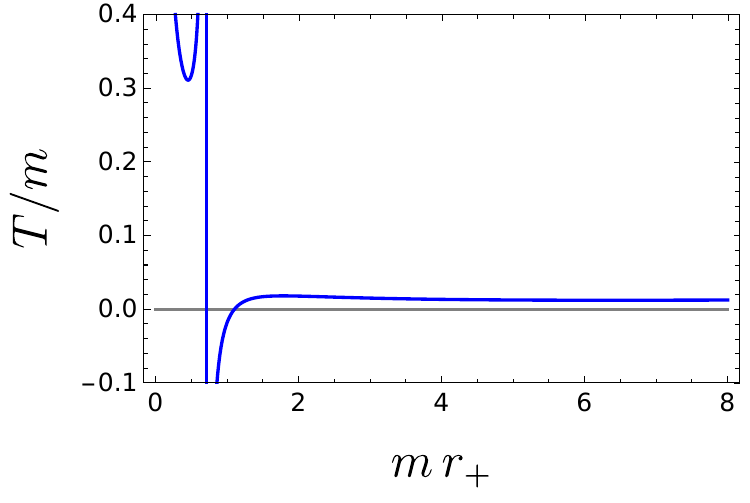}
	\includegraphics[width=0.496\textwidth]{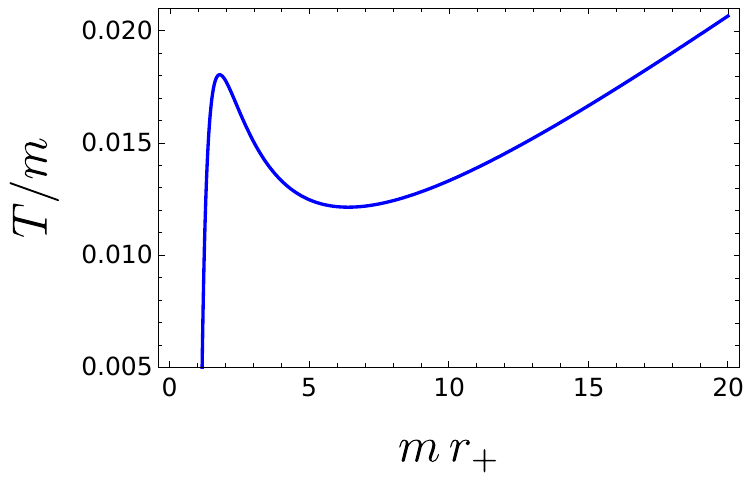}
	\includegraphics[width=0.496\textwidth]{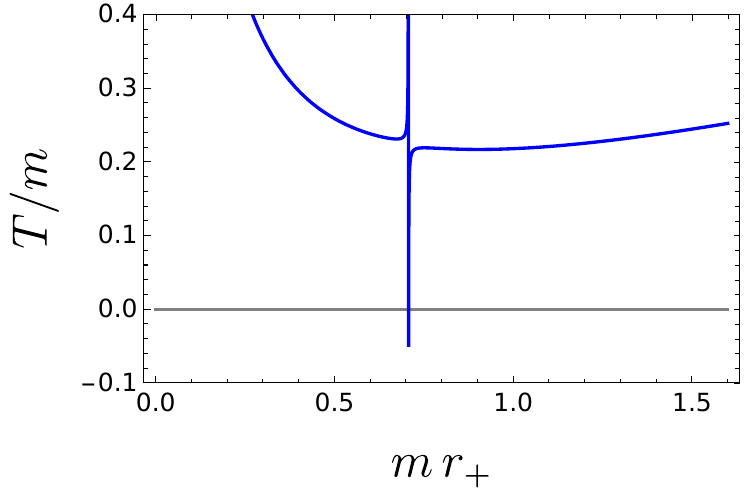}
	\includegraphics[width=0.496\textwidth]{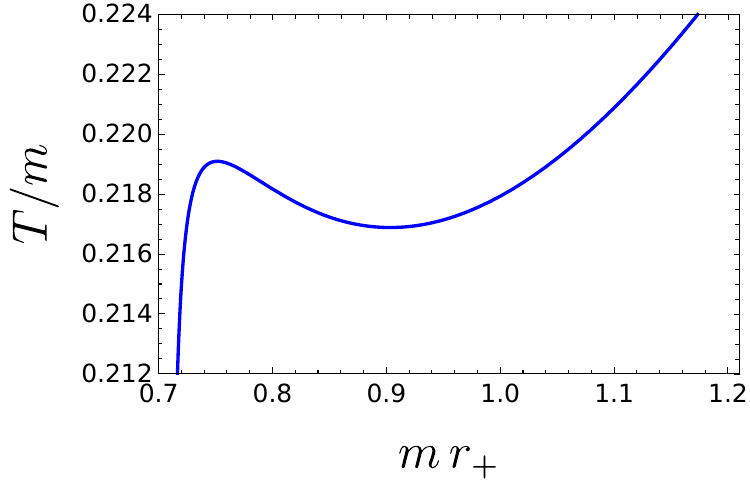}
	\caption{The temperature as a function of horizon radius for $k=-1$. \textit{Upper row}: We set $P=0.000466274 m^2 <P_c$ and plot from $0 \leq m r_+ < 8$ (left), with a close up (right) of the region to the right of the divergent point, where two local extrema appear. Note that the temperature diverges at $m r_+ =1/\sqrt{2}$ and is zero for $m r_+ =1.10244$. \textit{Lower row}: We set $\tilde{P}_c<P=0.0594566 m^2 <P_{max}$ and plot from $0 \leq m r_+ < 1.5$ (left), with a close up (right) of the region to the right of the divergent point, where again two local extrema appear. The temperature diverges at $m r_+ =1/\sqrt{2}$ and is zero for $m r_+ =0.707611$.
	}
	\label{fig:temp:crit}
\end{figure*}

\begin{figure}[htp]
	\centering
	\includegraphics[width=0.45\textwidth]{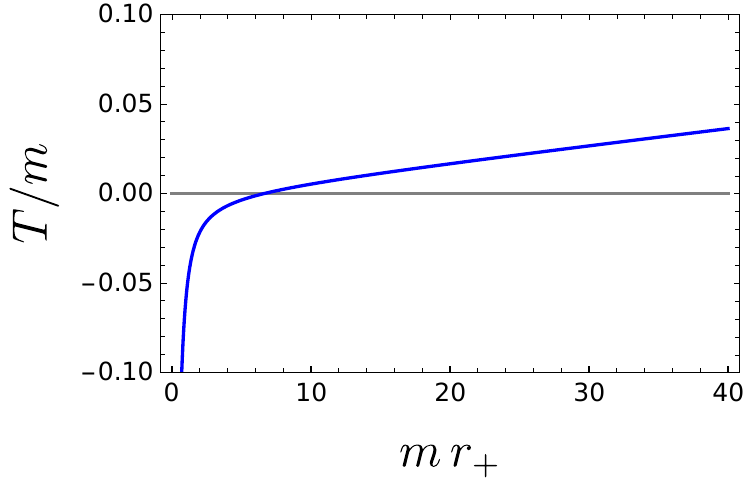}
	\includegraphics[width=0.45\textwidth]{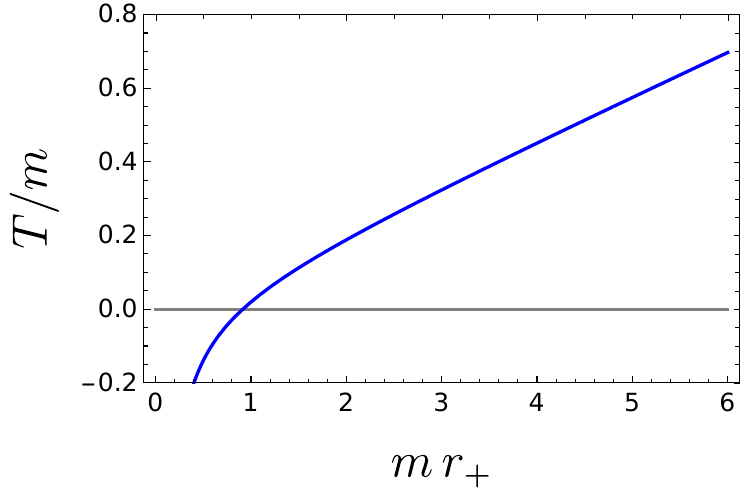}
	\caption{The temperature as a function of horizon radius for $k=1$. \textit{Top}: We set $P=0.000466274 m^2 <P_c$. The temperature is positive for $m r_+ > 6.62435$. \textit{Bottom}: We set $\tilde{P}_c<P=0.0594566 m^2 <P_{max}$. The temperature is positive for $m r_+ > 0.913957$.}
	\label{fig:temp:kp}
\end{figure}

In figure \ref{fig:temp:crit} we plot the temperature as a function of $r_+$ for the hyperbolic $k=-1$ case , with $P<P_c$ as well as $\tilde{P}_c<P<P_{max}$. In these cases we observe two local extrema in the large (cold) branch to the right of the divergent point, in addition to a local minimum at the left of the divergent point on the small (hot) branch. For comparison and later use, we have also plotted the temperature for $P<P_c$ and $\tilde{P}_c<P<P_{max}$ for the spherical $k=1$ case in figure \ref{fig:temp:kp}.

\section{Reverse isoperimetric inequality and thermodynamic instabilities} \label{sec:rii} 

The physical interpretation of thermodynamic volume is not clear. In simple cases such as Schwarzschild-AdS black holes, thermodynamic volume equals geometric volume, but in more general cases there is no known relation between them. For $k=0$ the thermodynamic volume \eqref{eqn:volume} reduces to
\be
V_{k=0}=\frac{512 \pi^2P r_+^3}{9\mu \kappa^2},
\ee
which is non-negative for $\mu>0$ and increases with pressure and the cube of the event horizon radius (recall from eq. \eqref{eqn:mass} that the coupling $\mu$ must be positive for the mass of the black hole to be positive).

\begin{figure}[htp]
	\centering
	\includegraphics[width=0.45\textwidth]{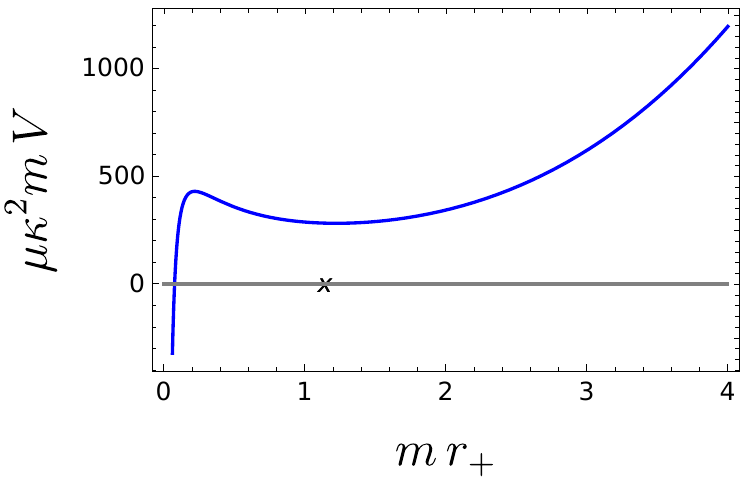}
	\includegraphics[width=0.45\textwidth]{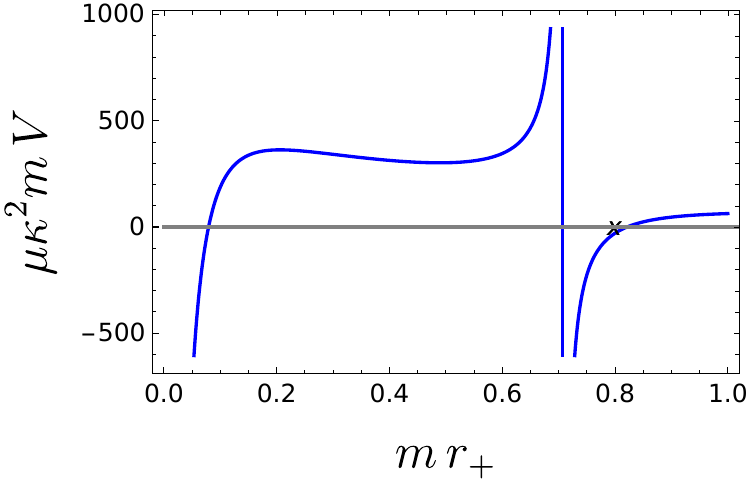}
	\caption{Thermodynamic volume as a function of horizon radius. The crosses show the point at which the temperature changes sign from negative to positive (see figure \ref{fig:temp}). \textit{Top}: the spherical $k=1$ case, \textit{bottom}: the hyperbolic $k=-1$ case. We have taken $P=0.0298416 m^2$ for both figures. The qualitative behavior does not depend on the specific value of $P$.}
	\label{fig:vol}
\end{figure}

In figure~\ref{fig:vol} we depict the thermodynamic volume for the $k=\pm 1$ cases as a function of the event horizon radius. In both cases the thermodynamic volume can be negative. For the spherical $k=1$ case, the thermodynamic volume is negative only in the unphysical region where the temperature is negative (see figure \ref{fig:temp}). However if $k=-1$ the thermodynamic volume is not only negative in the unphysical regions of phase space for the large (cold) black holes, but can also be negative in the physical region of the small (hot) black holes\footnote{We note that negativity of the thermodynamic volume also occurs for charged $z=3$ Ho\v{r}ava-Lifshitz black holes~\cite{poshteh2017} and for the Euclidean Taub-NUT solutions \cite{Johnson:2014xza}.}.
Like the temperature, the thermodynamic volume also diverges at $m r_+ = 1/\sqrt{2}$.

\subsection{Reverse isoperimetric inequality}

Here we consider the reverse isoperimetric inequality for $z=4$ Ho\v{r}ava-Lifshitz black holes. The reverse isoperimetric inequality 
\be\label{rpi}
\mathcal{R} = \bigg(\frac{(d-1)V}{\omega_{d-2}} \bigg)^{\frac{1}{d-1}}\Big(\frac{\omega_{d-2}}{A} \Big)^{\frac{1}{d-2}} \geq 1,
\ee
is a conjecture in $d$ dimensions, whose implication is that for a given thermodynamic volume, the entropy is maximized for Schwarzschild-AdS black holes~\cite{cvetivc2011}. This interpretation will not hold for the Ho\v{r}ava-Lifshitz case since the entropy \eqref{eqn:ent} is no longer proportional to the area of the black hole.

In $3+1$ dimensions we obtain
\be
\mathcal{R}=\frac{\sqrt[3]{3V}}{r_+}\geq 1,\label{eqn:rii}
\ee
where the horizon area $A=\omega_k r_+^2$~\cite{cai1999}. Using eq. (\ref{eqn:volume}) this becomes
\begin{widetext}
\begin{align}
	\mathcal{R}=&\sqrt[3]{\frac{32 \pi  \left(3 k^2+12 k m^2 r_+^2+64 \pi  m^2 P r_+^4\right)}{3 \kappa ^2 \mu  m^2 r_+^4}} \nn\\
	&\qquad\times\sqrt[3]{1+\frac{3 \left(5 k^2+4 k m^2 r_+^2-64 \pi  m^2 P r_+^4\right) \bigl\{3 k m^2+8 \pi  k P\left[1 + \ln \left(\frac{16}{3} \pi  P r_+^2\right)\right]+16 \pi  m^2 P r_+^2\bigr\}}{32 \pi  P
			\left(k+2 m^2 r_+^2\right) \left(3 k^2+12 k m^2 r_+^2+64 \pi  m^2 P r_+^4\right)}}, 
\end{align}
\end{widetext}
We find that eq. (\ref{eqn:rii}) can be violated for any $k=1,\, 0,\, -1$. For large black holes we find that 
\be \label{eqn:riik0}
\lim_{r_+\rightarrow\infty}\mathcal{R}_{k=\pm1} = 8\sqrt[3]{\frac{\pi^2 P}{3\mu \kappa^2}} = \mathcal{R}_{k=0},
\ee
where the latter equality holds for all values of $r_+$. For large black holes with $k=\pm1$ and for all sized black holes with $k=0$, the reverse isoperimetric inequality is violated for
\be
P<\bar{P}= \frac{3 \mu\kappa^2}{512\pi^2} = \frac{\mu\kappa^2}{32\pi m^2} P_{max}
,\label{eqn:pbar}
\ee
which is independent of $r_+$ and $m$, where we have used \eqref{eqn:press:ul}. We see that for $\mu\kappa^2/m^2>32\pi$ the upper bound $\bar{P}$ on the pressure is larger than $P_{max}$, and so the reverse isoperimetric inequality is always violated for the planar $k=0$ case, and the large $r_+$ limit for $k=\pm 1$.

\begin{figure}[htp]
	\centering
	\includegraphics[width=0.45\textwidth]{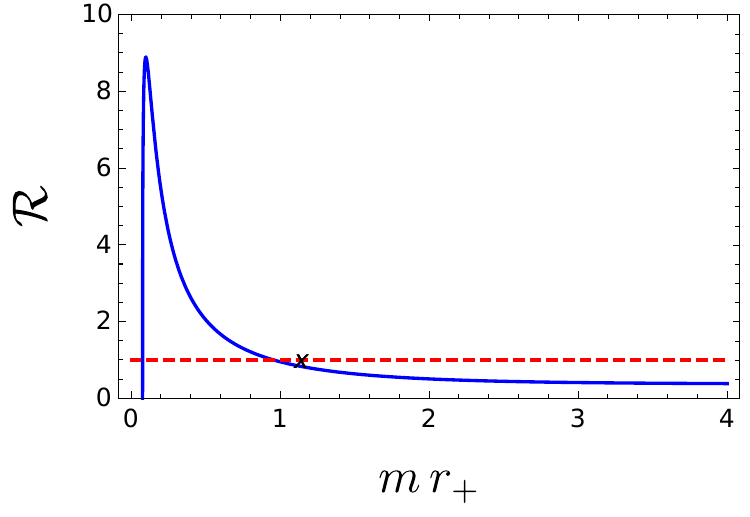}
	\includegraphics[width=0.45\textwidth]{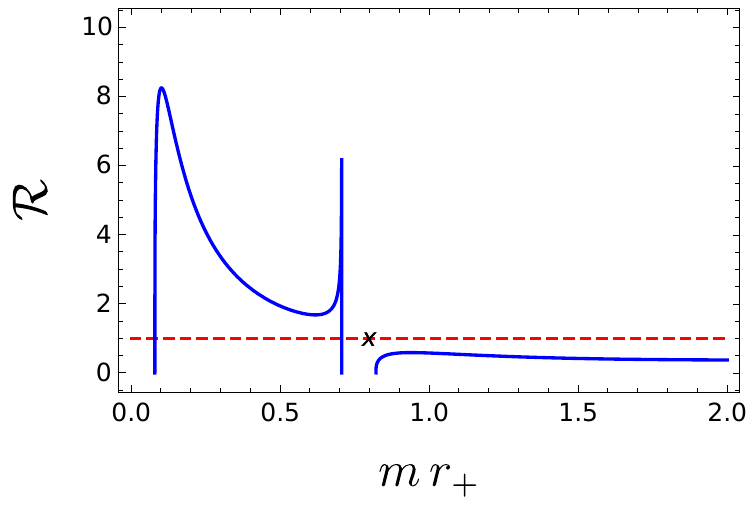}
	\caption{Isoperimetric ratio $\mathcal{R}$ as a function of the event horizon radius for $\mu\kappa^2/m^2=10^3>32\pi$. We have taken $P=0.0298416 m^2<P_{max}=3m^2/(16\pi)<\bar{P}=3000m^2/(512\pi^2)$. The dashed (red) line shows $\mathcal{R}=1$. The cross marks show the radius at which the temperature is zero. \textit{Top}: $k=1$ case. We find that $\mathcal{R}<1$ for $m r_+<0.0764301$ and $m r_+>0.954745$. Note that the black hole is unphysical in the region left of the cross mark, since the temperature is negative there (see figure \ref{fig:temp}).
	\textit{Bottom}: $k=-1$ case. $\mathcal{R}<1$ for $m r_+<0.0793416$. It diverges at $m r_+ = 1/\sqrt{2}$ which is the same point where the thermodynamic volume and the temperature diverge. The isoperimetric ratio is also less than one to the right hand side of the divergent point. We note that the temperature is negative in the region between the divergent point and the cross mark. Other regions are physical.}
	\label{fig:riiexpbar}
\end{figure}

In figure \ref{fig:riiexpbar} we plot $\mathcal{R}$ for $k=\pm1$ as a function of the event horizon radius for $\mu\kappa^2/m^2>32\pi$, implying $P<P_{max}<\bar{P}$. In these plots we have only shown regions where $\mathcal{R} > 0$, with the dashed line indicating $\mathcal{R} = 1$. We see that the reverse isoperimetric inequality is violated for small and large black holes in both cases. However for $k=1$, the violation for small black holes is in the unphysical region where the temperature is negative. For large black holes the inequality is violated in the physical region for both $k=\pm 1$, where both temperature and thermodynamic volume are positive. We note that for $k=\pm1$, the reverse isoperimetric inequality \eqref{eqn:rii} is also obviously violated in regions where the thermodynamic volume is negative.

\begin{figure}[htp]
	\centering
	\includegraphics[width=0.45\textwidth]{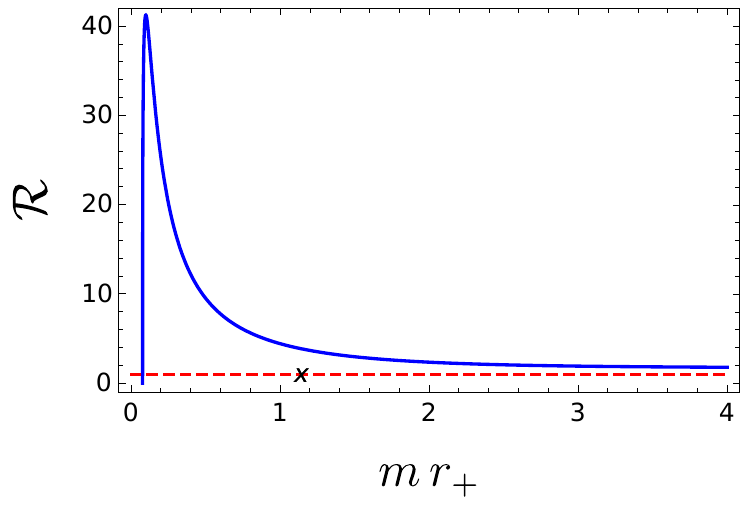}
	\includegraphics[width=0.45\textwidth]{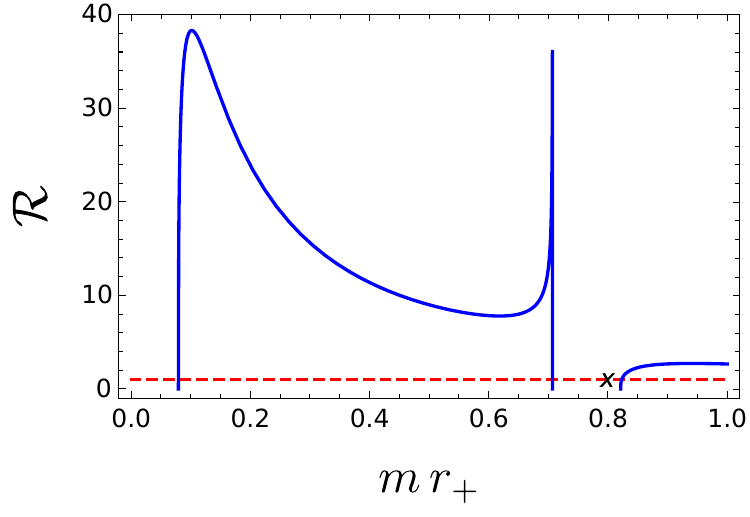}
	\caption{$\mathcal{R}$ as a function of the event horizon radius for $\mu\kappa^2/m^2=10<32\pi$. We have taken $\bar{P}=30m^2/(512\pi^2)<P=0.0298416 m^2<P_{max}=3m^2/(16\pi)$. The dashed (red) line shows $\mathcal{R}=1$. The cross marks show the radius at which the temperature is zero. \textit{Top}: $k=1$ case. We find that $\mathcal{R}<1$ for $m r_+<0.0764202$, which is in the unphysical region as the temperature is negative in the region left of the cross mark (see figure \ref{fig:temp}). \textit{Bottom}: $k=-1$ case. We see that $\mathcal{R}<1$ for $m r_+<0.0793289$ and $1/\sqrt{2}< m r_+ < 0.823114$; $\mathcal{R}$ diverges at $m r_+ = 1/\sqrt{2}$. Recall that the temperature is negative in the region between the divergent point and the cross mark. Other regions have positive temperature.}
	\label{fig:rii}
\end{figure}

If $\mu\kappa^2/m^2<32\pi$, there are two interesting cases: black holes for which $\bar{P}<P<P_{max}$ and those for which $P<\bar{P}<P_{max}$. We illustrate the former case in figure \ref{fig:rii}. Any violations of the reverse isoperimetric inequality for $k=1$ occur in unphysical regions where the temperature is negative. If $k=-1$ (the bottom panel of figure \ref{fig:rii}) we observe violations of the reverse isoperimetric inequality for small black holes; these are in the physical region where temperature is positive. The inequality is also violated for some intermediate size black holes. For large black holes there are no violations, as expected since $P>\bar{P}$. As $P\rightarrow \bar{P}$, we find $\mathcal{R}\rightarrow 1$ in the large $r_+$ limit.

\begin{figure*}[htp]
	\centering
	\includegraphics[width=0.496\textwidth]{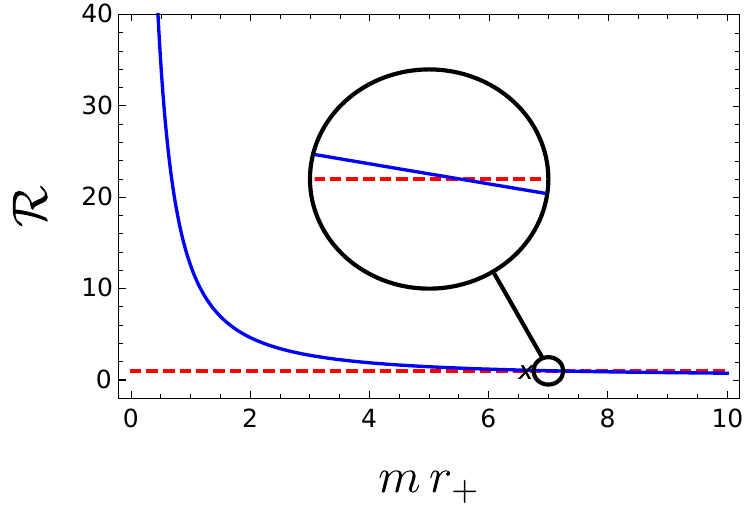}
	\includegraphics[width=0.496\textwidth]{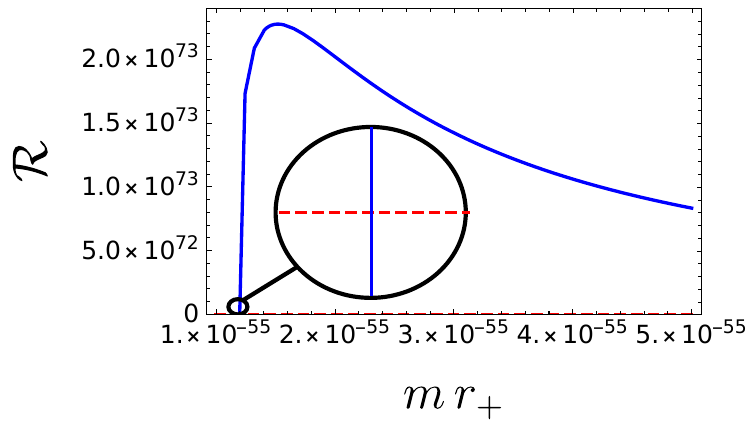}
	\includegraphics[width=0.496\textwidth]{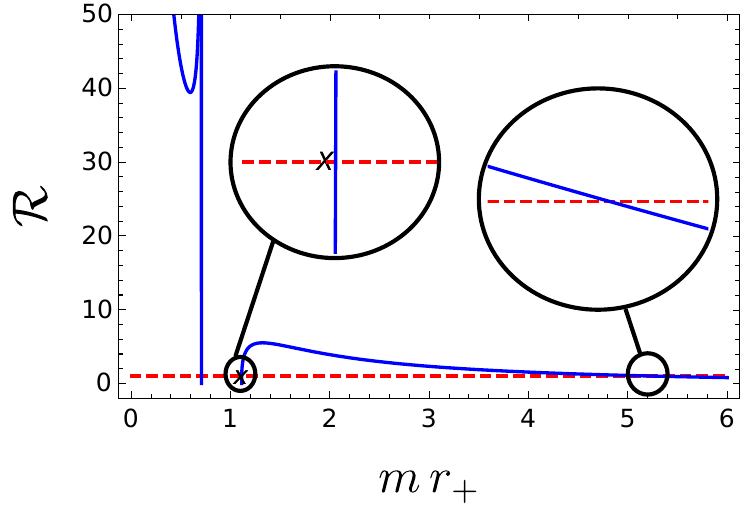}
	\includegraphics[width=0.496\textwidth]{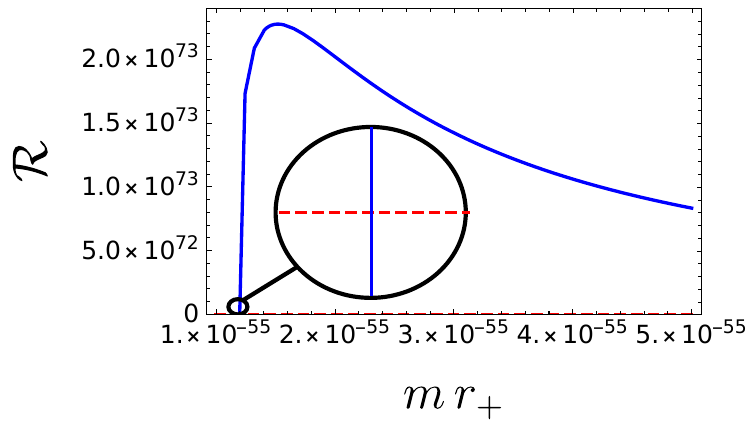}
	\caption{$\mathcal{R}$ as a function of the event horizon radius for $\mu\kappa^2/m^2=10<32\pi$. We have taken $P=0.000466274 m^2< \bar{P}=30m^2/(512\pi^2)<P_{max}=3m^2/(16\pi)$. The dashed (red) line shows $\mathcal{R}=1$. The crosses show the radius at which the temperature is zero. \textit{Upper row}: $k=1$ case. We plot from $0 \leq m r_+ < 10$ (left) and provide a close up (right) in a region of small black holes. The reverse isoperimetric inequality is violated for $m r_+ < 1.18317\times 10^{-55}$ and $m r_+> 7.11536$. Recall from figure \ref{fig:temp:kp} that for radii to the right of the cross mark the black holes are in the physical region with positive temperature. \textit{Lower row}: $k=-1$ case. We plot from $0 \leq m r_+ < 6$ (left) and provide a close up (right) in a region of small black holes. The reverse isoperimetric inequality is violated for $m r_+ < 1.18317\times 10^{-55}$, $1/\sqrt{2}< m r_+ < 1.11267$ and $m r_+> 5.28038$ ($m r_+ = 1/\sqrt{2}$ is the divergent point of $\mathcal{R}$). Note from figure \ref{fig:temp:crit} that the temperature is negative for black holes with radii between the divergent point and the cross mark, which here is at $m r_+ =1.10244$.}
	\label{fig:riisp}
\end{figure*}

We illustrate in figure \ref{fig:riisp} the situation for $P<\bar{P}<P_{max}$ and $\mu\kappa^2/m^2<32\pi$. For $k=1$, as before violations of the inequality \eqref{rpi} for very small black holes (see upper right diagram in figure \ref{fig:riisp}) are in the unphysical negative temperature regime. However at positive temperatures, sufficiently large black holes will violate \eqref{rpi}, shown in the inset in the upper left diagram of figure \ref{fig:riisp}. For $k=-1$, the reverse isoperimetric inequality is violated in three different regions: very small $r_+$, intermediate $r_+$, and large $r_+$. Note in the lower left diagram of figure \ref{fig:riisp} that the black holes with radius between the divergent point and cross mark are unphysical with negative temperature.
In other parts the temperature is positive; hence we have physical black holes violating the reverse isoperimetric inequality. Note also from the right diagrams of figure \ref{fig:riisp} that for small black holes the isoperimetric ratio behaves very similar in cases $k=\pm 1$.

\subsection{Thermodynamic instabilities}

We have seen that for $z=4$ Ho\v{r}ava-Lifshitz gravity, there are black holes that violate the reverse isoperimetric inequality \eqref{rpi}. Such violations have been noted before \eqref{eqn:rii}~\cite{hennigar2015}, with the associated black holes being called super-entropic, since their entropy exceeds the maximum implied from the relation \eqref{rpi} (see also~\cite{Boudet:2020eyr}). However for the black holes we are considering entropy is not proportional to area; we shall therefore refer to black holes violating \eqref{rpi} as
RII-violating black holes.

It was recently conjectured~\cite{johnson2020} that for super-entropic black holes the specific heat at constant volume, $C_V$, is negative. This conjecture was subsequently followed up with a counterexample~\cite{cong2019}, suggesting a broader version of the conjecture: 
black holes violating the reverse isoperimetric inequality \eqref{eqn:rii} are thermodynamically unstable. In specific terms, for such black holes, the specific heat at constant pressure, $C_P$ is negative whenever $C_V>0$. 

We consider this conjecture for the $z=4$ Ho\v{r}ava-Lifshitz RII-violating black holes. The specific heat at constant pressure and volume are
\begin{widetext}
\begin{eqnarray}
	C_P=T\left.\frac{\partial S}{\partial T}\right|_{P}&=&\frac{4 \pi  \left(k+2 m^2 r_+^2\right)^2 \left(-5 k^2-4 k m^2 r_+^2+64 \pi  m^2 P r_+^4\right)\left(3 k^2+12 k m^2 r_+^2+64 \pi  m^2 P r_+^4\right)}{\left(3 \mu\kappa^2 m^4 r_+^4\right)\left[5 k^3+26 k^2 m^2 r_+^2+8 k m^2 r_+^4 \left(m^2+24 \pi  P\right)+128 \pi  m^4 P r_+^6\right]}, \label{eqn:cp}\\
	C_V=T\left.\frac{\partial S}{\partial T}\right|_{V}&=&T\left(\left.\frac{\partial S}{\partial T}\right|_{P}+\left.\frac{\partial S}{\partial P}\right|_{T}\left.\frac{\partial P}{\partial T}\right|_{V}\right)=\frac{4 \pi  \left(-5 k^2-4 k m^2 r_+^2+64 \pi  m^2 P r_+^4\right)}{3 \mu\kappa^2 m^4 r_+^4 } \nn\\
	&\qquad&\qquad\qquad\qquad \times\left[\frac{\left(k+2 m^2 r_+^2\right)^2 \left(3 k^2+12 k m^2 r_+^2+64 \pi  m^2 P r_+^4\right)}{\left(5 k^3+26 k^2 m^2 r_+^2+8 k m^2 r_+^4 \left(m^2+24 \pi  P\right)+128 \pi  m^4 P r_+^6\right)}-\frac{I}{J}\right], \label{eqn:cv}
\end{eqnarray}
where
\begin{eqnarray}
	I&=&12 m^2 r_+^4 \left\{k^4
	\left(15 m^2-8 \pi  P\right)+6 k^3 m^2 r_+^2 \left(13 m^2-16 \pi  P\right)+8 k^2 m^2 r_+^4 \left(3 m^4+12 \pi  m^2 P+64 \pi ^2 P^2\right)+8 \pi  k P\right. \nn\\
	&\times&\ln \left(\frac{16}{3} \pi 
	P r_+^2\right) \left[5 k^3+26 k^2 m^2 r_+^2+8 k m^2 r_+^4 \left(m^2+24 \pi  P\right)+128 \pi  m^4 P r_+^6\right]
	-\left.256 \pi  k m^6 P r_+^6-2048 \pi ^2 m^6 P^2 r_+^8\right\}^2, \nn\\
	J&=&75 k^6 \left(3 m^2-8 \pi  P\right)+450 k^5 m^2 r_+^2 \left(3 m^2-8 \pi  P\right)+16 k^4 m^2 r_+^4 \left(81 m^4+144 \pi  m^2 P-1024 \pi ^2 P^2\right) + 96 k^3 m^4 r_+^6\nn\\
	&\times& \left(m^2-8 \pi  P\right) \left(3 m^2-8 \pi  P\right)-49152 \pi ^2 k^2 m^4 P^2 r_+^8 \left(m^2-2 \pi  P\right)+8192 \pi ^2 k m^6 P^2 r_+^{10} \left(5 m^2-8 \pi  P\right)+262144 \pi ^3 m^8 P^3 r_+^{12}. \nn
\end{eqnarray}
\end{widetext}

In the $k=0$ flat horizon case, the specific heats \eqref{eqn:cp} and \eqref{eqn:cv} reduce to
\ba
C_{P,k=0}&=&\frac{512 \pi ^2 P r_+^2}{3\mu\kappa^2}, \nn\\ C_{V,k=0}&=&\frac{128 \pi ^2 P r^2}{3 \mu\kappa^2},
\ea
which are always positive. Since for $k=0$, the reverse isoperimetric inequality is violated for $P<\bar{P}=3 \mu\kappa^2/(512\pi^2)$ (see \eqref{eqn:riik0}),
we have a counterexample for the aforementioned conjectures~\cite{johnson2020,cong2019}.

\begin{figure*}[htp]
	\centering
	\includegraphics[width=0.496\textwidth]{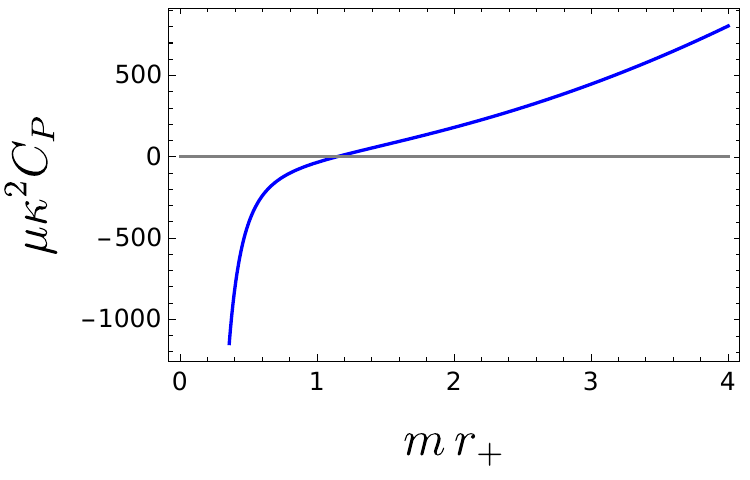}
	\includegraphics[width=0.496\textwidth]{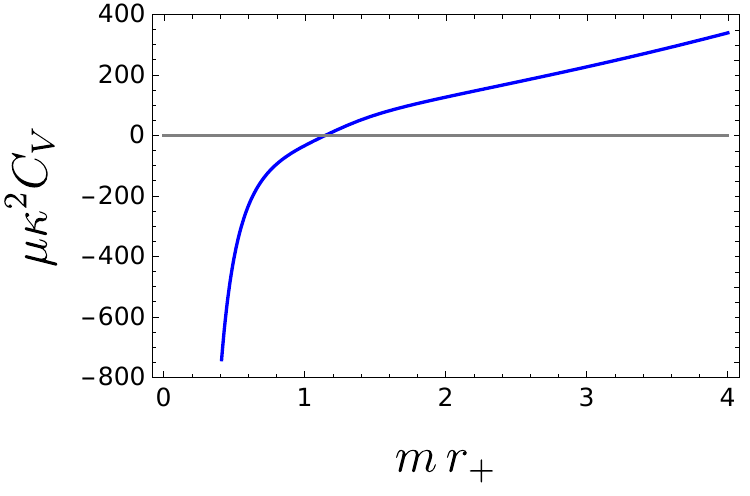}
	\includegraphics[width=0.496\textwidth]{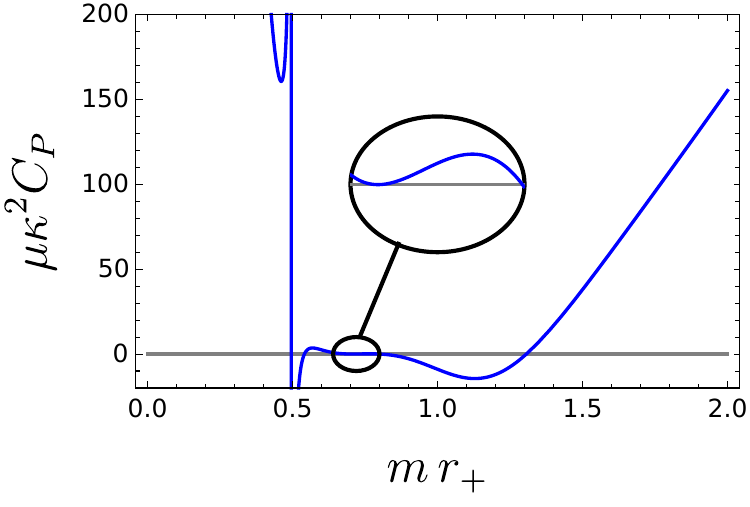}
	\includegraphics[width=0.496\textwidth]{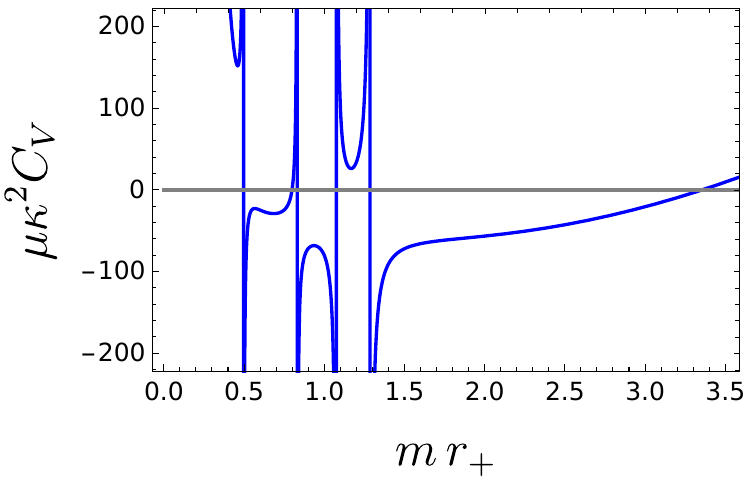}
	\caption{ The specific heats as a function of horizon radius for $P=0.0298416 m^2$. \textit{Upper row}: $k=1$ case. The specific heats $C_P$ and $C_V$ are both positive (negative) in the region that the temperature is positive (negative). Both $C_P$ and $C_V$ vanish at $m r_+ = 1.14244$, where the temperature is zero.
		\textit{Lower row}: $k=-1$ case. Both $C_P$ and $C_V$ are positive for $m r_+ < 0.0793416$. In the region $0.799057<m r_+<0.823114$ the specific heat $C_P$ is negative but $C_V$ is positive. We note that $C_P$ and $C_V$ are both positive for $m r_+ > 3.35605$. Note also that the specific heat at constant pressure $C_P$ diverges at the point where the temperature is minimum (see figure \ref{fig:temp}). It also has four zeros, related to the minimum of the entropy, the maximum of the entropy (where the temperature also diverges), a zero of the temperature, and another minimum of the entropy, respectively from left to right.}
	\label{fig:c}
\end{figure*}

\begin{figure*}[htp]
	\centering
	\includegraphics[width=0.496\textwidth]{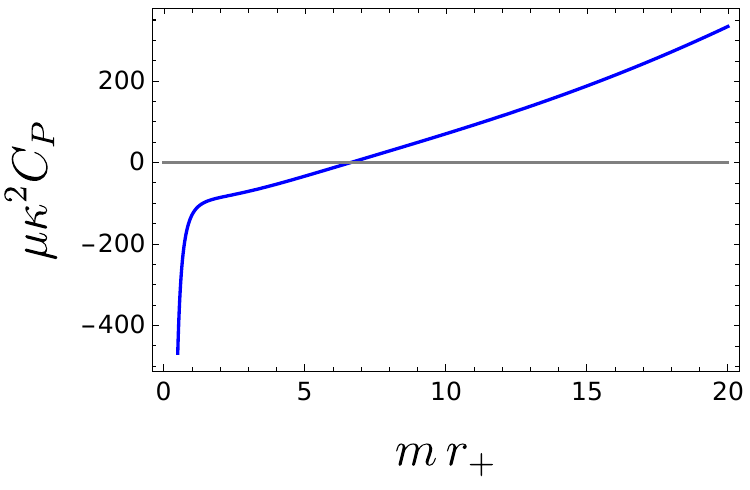}
	\includegraphics[width=0.496\textwidth]{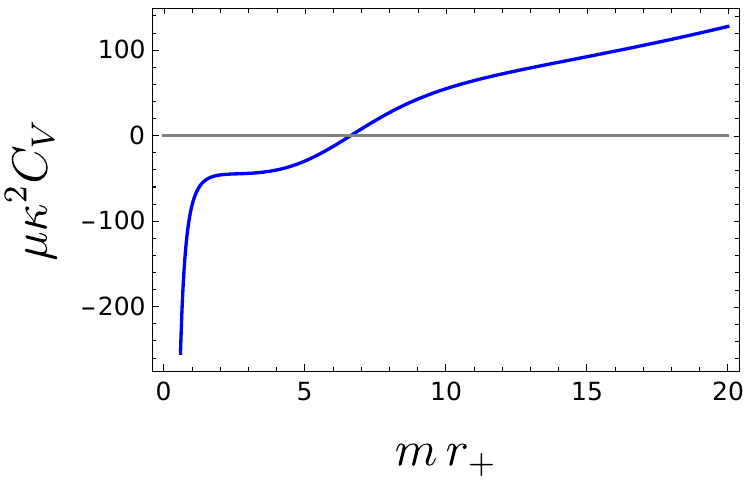}
	\includegraphics[width=0.496\textwidth]{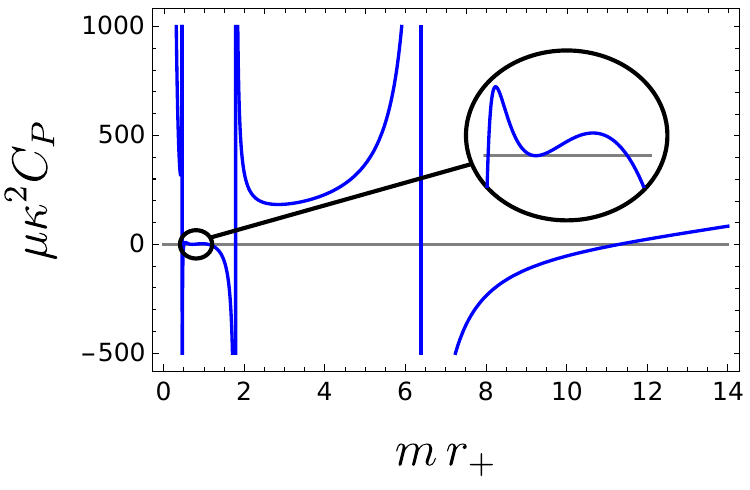}
	\includegraphics[width=0.496\textwidth]{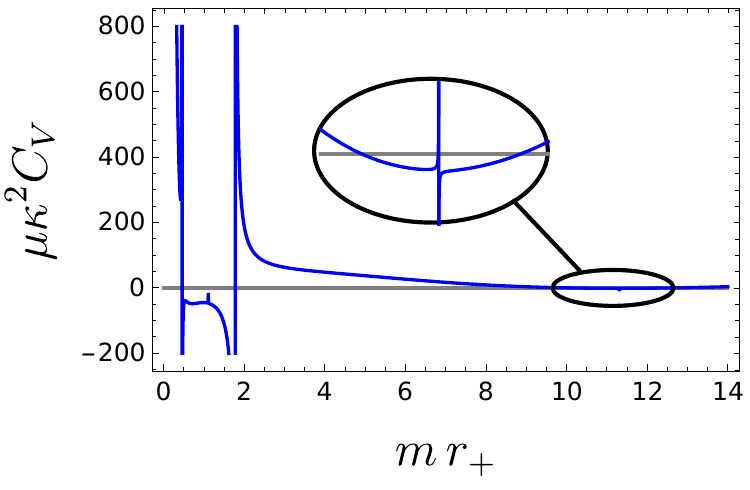}
	\caption{ The specific heats as a function of horizon radius for $P=0.000466274 m^2$. \textit{Upper row}: $k=1$ case. The specific heats $C_P$ and $C_V$ are both positive (negative) in the region that the temperature is positive (negative). As for the temperature, the zero of $C_P$ and $C_V$ is at $m r_+ = 6.62435$ (see the top panel of figure \ref{fig:temp:kp}). \textit{Lower row}: $k=-1$ case. Both $C_P$ and $C_V$ are positive for $m r_+ < 1.18317\times 10^{-55}$. In the region $1.10244<m r_+<1.11267$ both $C_P$ and $C_V$ are negative, and for for $m r_+ > 12.5723$ $C_P$ and $C_V$ are both positive. We also note that $C_P$ diverges at three points, corresponding to extrema of the temperature (see the upper row of figure \ref{fig:temp:crit}). It also has four zeros related to a minimum of the entropy, the maximum of the entropy (where the temperature also diverges), a zero of the temperature, and another minimum of the entropy, respectively from left to right.}
	\label{fig:csp}
\end{figure*}

To have a comparison with figures~\ref{fig:riiexpbar}, \ref{fig:rii} and \ref{fig:riisp} we plot the specific heat for cases $k=\pm 1$ for two values of pressure in figures~\ref{fig:c} and~\ref{fig:csp}. We are only interested in the signs of the specific heats $C_P$ and $C_V$. The quantity $\mu\kappa^2$ is a positive factor in eqs. \eqref{eqn:cp} and \eqref{eqn:cv} and its specific value does not change the sign of $C_P$ or $C_V$. Hence it is also not relevant to specify the value of $\bar{P}$ in analyzing $C_P$ and $C_V$ (see eq.~\eqref{eqn:pbar}).

We saw in the top panel of figure~\ref{fig:riiexpbar} that, for the case of spherical horizon $k=1$ and $\mu\kappa^2/m^2>32\pi$, the reverse isoperimetric inequality is violated for large black holes. These black holes are in the physical region where the temperature is positive. Now from the upper row of figure~\ref{fig:c} we see that for these black holes both $C_P$ and $C_V$ are positive. Therefore, in this case, the $z=4$ Ho\v{r}ava-Lifshitz RII-violating black holes are thermodynamically stable.

Consider next the bottom panel of figure~\ref{fig:riiexpbar} with $k=-1$ and $\mu\kappa^2/m^2>32\pi$. By comparing this plot with the lower row of figure~\ref{fig:c} we see that the large $z=4$ Ho\v{r}ava-Lifshitz RII-violating black holes are also thermodynamically stable in this case. For the particular choice $P=0.0298416 m^2$ of the pressure we have taken in figures~\ref{fig:riiexpbar} and \ref{fig:c}, these large black holes are in the region with $m r_+ > 3.35605$, which is in the physical region with positive temperature. We also note that there are some small black holes ($m r_+ < 0.0793416$), with hyperbolic horizons and $\mu\kappa^2/m^2>32\pi$, which are RII-violating and thermodynamically stable.

Let us now consider the case of $\mu\kappa^2/m^2<32\pi$ and $\bar{P}<P<P_{max}$. From the top plot of figure~\ref{fig:rii} we see that for $k=1$ the physical black holes always satisfy the reverse isoperimetric inequality. We find from the upper row of figure~\ref{fig:c} that both $C_P$ and $C_V$ are positive for these black holes, so they are thermodynamically stable. For $k=-1$ hyperbolic horizons, we see from the bottom panel of figure~\ref{fig:rii} that the reverse isoperimetric inequality is violated in two physical regions: small black holes (with $m r_+<0.0793289$) and intermediate size black holes (with $0.799057<m r_+<0.823114$). We see in the lower row of figure~\ref{fig:c} that for the small black holes both $C_P$ and $C_V$ are positive. However, for the intermediate size black holes $C_P$ is negative but $C_V$ is positive.

Still considering $\mu\kappa^2/m^2<32\pi$ and $P<\bar{P}<P_{max}$, we now investigate the physical region with positive temperature. In the upper row of figure~\ref{fig:riisp} we saw that for $k=1$ spherical horizons, large black holes violate the reverse isoperimetric inequality. For the value of the pressure we have taken, $P=0.000466274 m^2$, this occurs for $m r_+> 7.11536$. Comparison with the upper row of figure~\ref{fig:csp} shows that for these black holes both $C_P$ and $C_V$ are positive. For $k=-1$ hyperbolic horizons, we saw in the lower row of figure~\ref{fig:riisp} that the reverse isoperimetric inequality is violated for some small, intermediate, and large black holes. We see from the lower row of figure~\ref{fig:riisp} that the small black holes are thermodynamically stable, wherears the intermediate size ones are thermodynamically unstable (with both $C_P$ and $C_V$ negative). Large black holes with $m r_+ > 12.5723$ have both $C_P$ and $C_V$ positive.

Over all, we find that for $P<\bar{P}$, the reverse isoperimetric inequality is violated for large black holes with $k=1$. These black holes are thermodynamically stable. If $k=-1$, the reverse isoperimetric inequality is violated not only for large black holes, but also for small ones, and in both cases we have thermodynamically stable black holes. Here if $\mu\kappa^2/m^2<32\pi$, we also have intermediate size RII-violating black holes. For these ones both $C_P$ and $C_V$ are negative.

If $P>\bar{P}$, (physical) black holes with spherical horizons ($k=1$) always satisfy the reverse isoperimetric inequality. These black holes are also thermodynamically stable. For hyperbolic horizons ($k=-1$), we have small and intermediate size RII-violating black holes. The small ones are thermodynamically stable but the intermediate ones have $C_P<0$ and $C_V>0$. Large $k=-1$ black holes also satisfy the reverse isoperimetric inequality and are thermodynamically stable.

\section{Critical behavior} \label{sec:crit}

In this section we investigate the critical behavior of $z=4$ Ho\v{r}ava-Lifshitz black holes. To find the equation of state we use eq. (\ref{eqn:temp}) to obtain
\be\label{eqn:eos}
P=\frac{T}{v}+\frac{k}{4 \pi v^2}+\frac{2 k T}{m^2 v^3}+\frac{5 k^2}{4 \pi m^2 v^4},
\ee
where we take $v=2r_+$ as the specific volume so that the above relation is similar to the equation of state of the Van der Waals system.

Contrary to typical behaviour in Einstein gravity \cite{kubizvnak2012,Kubiznak:2016qmn}, the $k=1$ case has no interesting phase behaviour or critical behaviour. 
Critical behavior, if any, takes place at an inflection point of the equation of state (\ref{eqn:eos}), namely
\be\label{eqn:inflection}
\frac{\partial P}{\partial v}=0, \qquad \frac{\partial^2 P}{\partial v^2}=0 \; .
\ee
No real positive solutions to both of the above equations exist for $k=1,\,0$, and so there is no critical behavior for black holes with spherical or planar horizons.

However for hyperbolic horizons with $k=-1$ we find two positive solutions 
\ba
v_{c}&=&\sqrt{\frac{2}{m^2} \left(6+\sqrt{21}\right)}, \nn\\ \tilde{v}_{c}&=&\sqrt{\frac{2}{m^2} \left(6-\sqrt{21}\right)} ,
\label{eqn:crit:2rad}
\ea
to \eqref{eqn:inflection}. 
Hence, in contrast to AdS black holes in general relativity \cite{Kubiznak:2016qmn}, critical behavior in $z=4$ Ho\v{r}ava-Lifshitz black holes takes place only for $k=-1$. In the rest of this section we focus on this case. We note that the critical points \eqref{eqn:crit:2rad} are the same critical points \eqref{eqn:crit:rad} at which the temperature has an inflection point; this is a consequence of temperature and pressure being linearly related.

Two points are in order here. In general relativity critical behavior occurs only in the $k=1$ case~\cite{kubizvnak2012}. In fact, by investigating the behavior of the temperature and the specific heat, it has been shown that there is a duality between topological black holes in general relativity with $k=-1,\, 0,\, 1$ and those in $z=3$ Ho\v{r}ava-Lifshitz gravity with $k=1,\, 0,\, -1$, respectively~\cite{cai2009:1,poshteh2017}. We see that a similar duality also exists for $z=4$. 

However there is a strong distinction between the two cases. There is no critical behavior in $z=3$ Ho\v{r}ava-Lifshitz black holes if the detailed balanced condition holds~\cite{mo2015}. This can be shown by taking $m\rightarrow\infty$ in eq. \eqref{eqn:eos}. But in Ho\v{r}ava-Lifshitz gravity with the detailed balanced condition, we find for $z=4$ that critical behaviour is present\footnote{It is interesting that by relaxing the detailed balanced condition, critical behaviour appears in $z=3$ Ho\v{r}ava-Lifshitz black holes with spherical horizons ($k=1$)~\cite{ma2017}.}.

\begin{figure}[htp]
	\centering
	\includegraphics[width=0.45\textwidth]{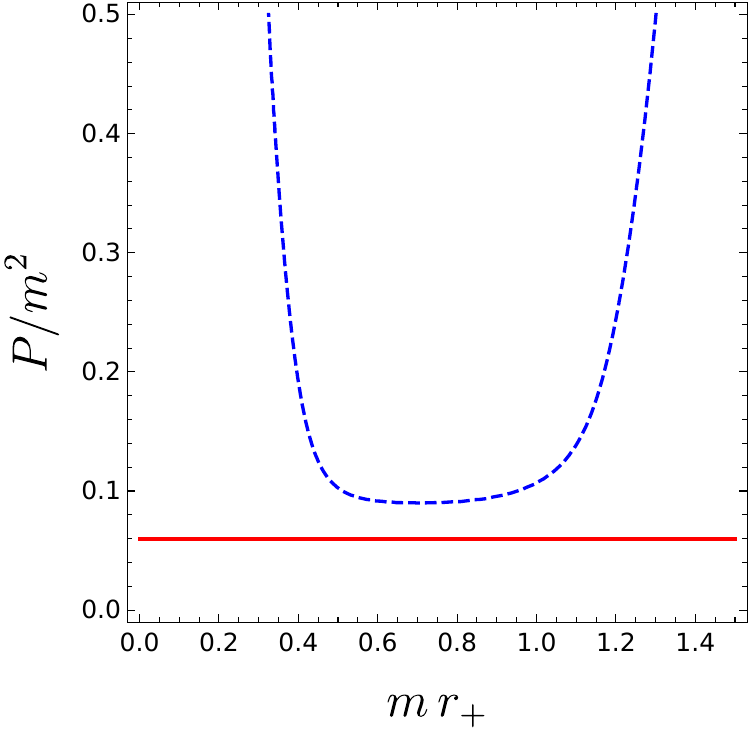}
	\caption{ The entropy is negative for the points above the dashed (blue) curve and positive below it. The solid (red) line separates the parameter space into a region for which the metric function is well behaved everywhere (below the line, where $P < P_{max}$), and a region for which the metric function is not real at large distances (above the line, $P > P_{max}$). We will work in the region below the solid (red) line which also guarantees that the entropy is positive. We have set $k=-1$.}
	\label{fig:cond}
\end{figure}

Before investigating the critical behaviour at $v_{c}$ and $\tilde{v}_{c}$, we examine the restrictions imposed on the thermodynamic variables. A first restriction comes from \eqref{eqn:press:ul}. Even though there are regions in parameter space where the entropy could be negative, as long as $P\leq P_{max}$ the entropy (\ref{eqn:ent}) is positive. We show this in figure~\ref{fig:cond}, where vanishing entropy corresponds to the dashed (blue) curve; above it the entropy is negative. The solid (red) line is the $P=3m^2/(16 \pi)$ surface: any point above it violates the condition \eqref{eqn:press:ul} (or equivalently \eqref{eqn:lambdacon}). Since the two curves do not touch, once (\ref{eqn:press:ul}) is satisfied, the entropy is positive.

\begin{figure}[htp]
	\centering
	\includegraphics[width=0.45\textwidth]{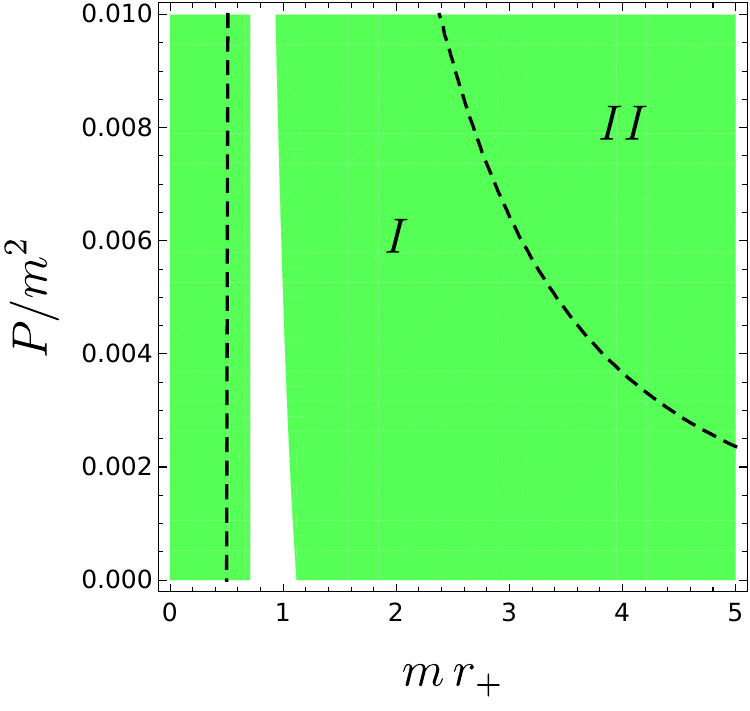}
	\caption{The shaded (green) parts show the (physical) regions where the temperature (\ref{eqn:temp}) is positive. The two dashed curves show the zeros of the mass (\ref{eqn:mass}), and $v_c$ and $\tilde{v}_c$ lie in regions $I$ and $II$, respectively. The divergent point of the temperature lies on the left hand side of region $I$. The regions continue from right and top, however we are restricted to $P\leq3m^2/(16 \pi)$. We have set $k=-1$.}
	\label{fig:region}
\end{figure}

Considering next eq. (\ref{eqn:mass}), we see that the mass cannot be negative (note that we are taking $\mu>0$), but for $k=-1$ it might be zero. Furthermore, as we have seen in the previous section, the temperature (\ref{eqn:temp}) can be negative in some regions of parameter space. Dismissing such regions as unphysical, in figure~\ref{fig:region} we plot the pressure as a function of the horizon radius, shading (in green) the regions for which the temperature is positive. The two dashed curves represent the points for which $M=0$; $v_c$ and $\tilde{v}_c$ lie in regions $I$ and $II$ respectively. The temperature is divergent on the left boundary of the white (unshaded) region.

\subsection{Criticality at $v_c$}

We consider first the behavior of our thermodynamic system near $v_c$. Using eqs. (\ref{eqn:eos}) and (\ref{eqn:inflection}) with $k=-1$, we find the critical values of temperature and pressure:
\ba
T_c&=&\frac{\sqrt{\left(33-7 \sqrt{21}\right)m^2}}{12 \pi }, \nn\\ P_c&=&\frac{\left(69-14 \sqrt{21}\right) m^2}{720 \pi }, \label{eqn:crittp}
\ea
which were also found in eqs. \eqref{eqn:crit:pres} and \eqref{eqn:crit:tem}.

\begin{figure}[htp]
	\centering
	\includegraphics[width=0.45\textwidth]{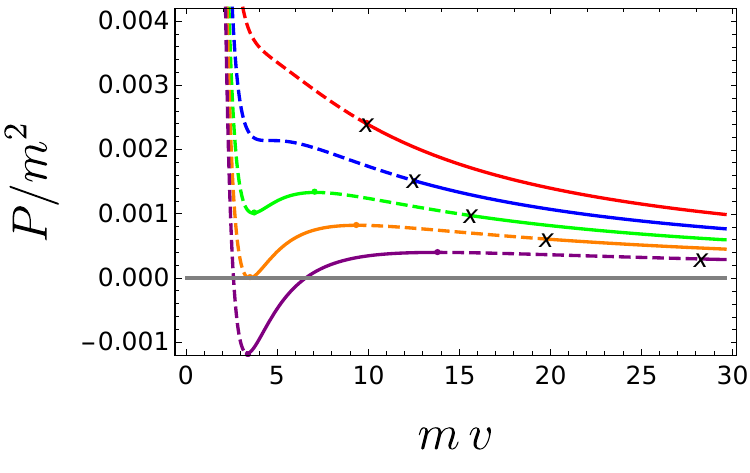}
	\includegraphics[width=0.45\textwidth]{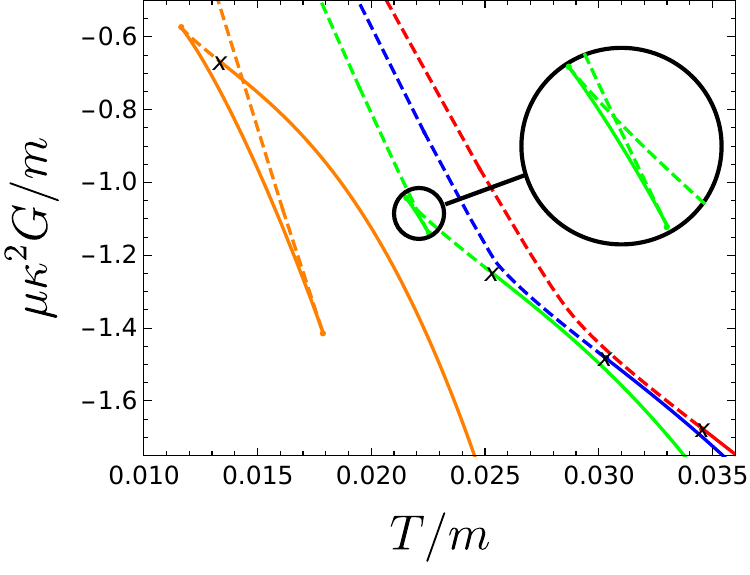}
	\caption{ Criticality at $v_c$ for hyperbolic horizons $k=-1$. The dashed parts of the plots show local instability as the specific heat $C_P$ is negative there. The crosses mark the points where the mass is zero. \textit{Top}: The $P-v$ diagrams are shown for $T=0.0320921 m>T_c$ (red), $T=T_c=0.0254699 m$ (blue), $T_0<T=0.0203759 m<T_c$ (green), $T=T_0=0.0160821 m$ (orange), and $T=0.0112575 m<T_0$ (purple), respectively from top to bottom. \textit{Bottom}: The $G-T$ plots for $P=1.3 P_c$ (red), $P=P_c=0.00214149 m^2$ (blue), $P=0.7 P_c$ (green), and $P=0.2 P_c$ (orange) respectively from right to left. Swallowtail behavior is observed for $P<P_c$.
	}
	\label{fig:crit}
\end{figure}

In the top panel of figure~\ref{fig:crit} we depict the $P-v$ diagrams near $v_c$. The isotherm at $T=T_c$ has an inflection point, and for $T<T_c$, have an oscillatory part. For temperatures below $T_0$, where
\be
T_0=\frac{1}{\pi}\sqrt{\frac{5m^2}{982+86 \sqrt{129}}},
\ee
the pressure is negative for some values of $v$. This violates the asymptotic structure that we have assumed for the spacetime, but such curves are valid where the equal area law can be applied.

The dashed parts of the curves in figure~\ref{fig:crit} are the regions in which the specific heat $C_P$ is negative, indicative of a local instability. The crosses on the curves show the points where the mass of the black hole is zero. On the left hand side of the crosses we are in region $I$ of figure~\ref{fig:region}, and on the right hand side of the crosses in region $II$. We see that for the temperatures and pressures that we have considered in figure~\ref{fig:crit}, the black holes of region $II$ are locally stable.

For the black holes whose parameters lie in region $I$ of figure~\ref{fig:region}, the specific heat is always negative as long as $T\geq T_c$. For $T<T_c$, intermediate size black holes in the oscillatory part are locally stable. This phenomenon is opposite to that which occurs in general relativity, where black holes in this region are unstable.

These peculiar features of $z=4$ Ho\v{r}ava-Lifshitz black holes are manifest in the free energy, shown in the bottom panel of figure~\ref{fig:crit}. We find an inverted swallowtail $P<P_c$: the first-order transition at the crossover is between large and small black holes, with an intermediate black hole, which is on a stable branch, corresponding to the warped part of the swallowtail curve. The small black hole is unstable. However, the large black hole can be locally stable if the pressure is small, as can be seen in the leftmost (orange) curve in the bottom panel of figure~\ref{fig:crit}. 

Recall that in general relativity $M=0$ results in $G=0$, corresponding to thermal AdS. Here we have a considerably different situation, where $M=0$ black holes can occur, shown by crosses in figure~\ref{fig:crit}. We take these crosses as the reference points --- which depend on the pressure (/cosmological constant). To the right of the crosses, each curve has a branch of large, high temperature, locally stable black holes in region $II$ (which are also globally preferred). To the left, for lower temperatures, the black holes are locally unstable for $P\geq P_c$. For $P<P_c$ the small and large black holes are locally unstable.

\subsection{Criticality at $\tilde{v}_c$}

Inserting $\tilde{v}_c$ and $k=-1$ into (\ref{eqn:eos}) and (\ref{eqn:inflection}), we find the temperature and the pressure 
\ba
\tilde{T}_c&=&\frac{\sqrt{\left(33+7 \sqrt{21}\right)m^2}}{12 \pi }, \nn\\ \tilde{P}_c&=&\frac{\left(69+14 \sqrt{21}\right) m^2}{720 \pi },
\ea
associated with $\tilde{v}_c$, which were also found in eqs. \eqref{eqn:crit:pres} and \eqref{eqn:crit:tem}. Comparing the above values with \eqref{eqn:crittp} we see that $\tilde{T}_c>T_c$ and $\tilde{P}_c>P_c$, so criticality at $\tilde{v}_c$ occurs for higher temperature and higher pressure. 

\begin{figure}[htp]
	\centering
	\includegraphics[width=0.45\textwidth]{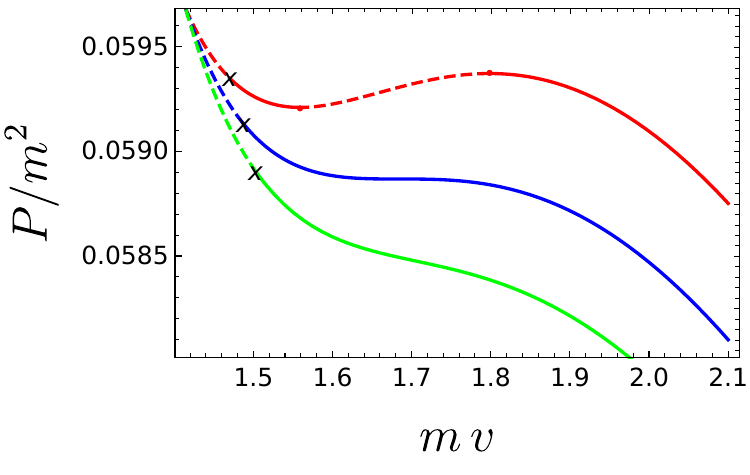}
	\includegraphics[width=0.45\textwidth]{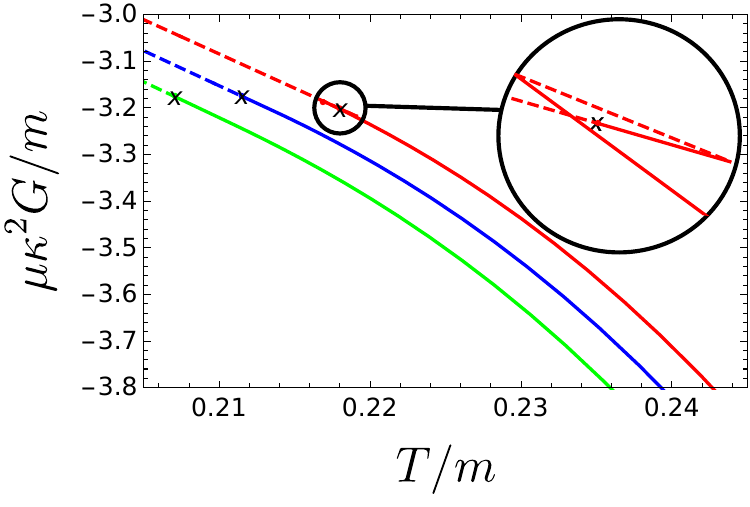}
	\caption{Criticality at $\tilde{v}_c$ for hyperbolic horizons $k=-1$. The dashed parts of the plots show local instability as the specific heat $C_P$ is negative there. The crosses mark the points where the mass is zero. \textit{Top}: The $P-v$ diagrams are shown for $T=0.216490 m>\tilde{T}_c$ (red), $T=\tilde{T}_c=0.213986 m$ (blue), and $T=0.211846 m<\tilde{T}_c$ (green), respectively from top to bottom. \textit{Bottom}: The $G-T$ plots for $P=0.0594566 m^2>\tilde{P}_c$ (red), $P=\tilde{P}_c=0.0588679 m^2$ (blue), and $P=0.0582792 m^2<\tilde{P}_c$ (green), respectively from right to left. Swallowtail behavior is observed for $P>\tilde{P}_c$. The inset shows swallowtail behavior only schematically -- the actual slopes are very similar, and would necessitate magnification beyond that which is feasible to illustrate.
	}
	\label{fig:tilde}
\end{figure}

The critical behavior at $\tilde{v}_c$ is shown in figure~\ref{fig:tilde}. The $P-v$ diagrams are shown on the left panel. The inflection point takes place at the $T=\tilde{T}_c$ isotherm. The crosses show the points where the mass vanishes, separating region $I$ from region $II$ (see figure~\ref{fig:region}). Since the inflection point is on the right hand side of the cross, this critical point is for black holes in region $II$. In contrast to the behavior near $v_c$, where the oscillatory part appears for black holes with $T<T_c$, we have an oscillatory part for isotherms with $T>\tilde{T}_c$. This reverse Van der Waals transition was first seen for hyperbolic black holes in Lovelock gravity \cite{frassino2014,Kubiznak:2016qmn}.
We also find that black holes in region $II$ are thermodynamically stable as long as their temperature is below $\tilde{T}_c$. For $T>\tilde{T}_c$ we have unstable black holes in the oscillatory part.

The $G-T$ plots are also presented in figure~\ref{fig:tilde} (the bottom panel). We find that swallowtail behavior appears for $P>\tilde{P}_c$, unlike criticality near $v_c$ where swallowtail behavior is observed $P < P_c$. Intermediate sized black holes, on the warped branch of the swallowtail, are thermodynamically unstable. As temperature increases, there is a first-order phase transition from a small (unstable) black hole with negative specific heat to a large black hole with positive specific heat, shown in the inset in the top panel of figure~\ref{fig:tilde}.

\section{Concluding remarks}\label{sec:con}

We have shown that $z=4$ Ho\v{r}ava-Lifshitz black holes have a rich thermodynamic structure, with a number of surprising features. They exhibit both Van der Waals and reverse Van der Waals behaviour, but with notably distinct stability features. The Van der Waals behaviour is characterized by an inverted swallowtail structure in the free energy, where the first-order phase transition is between stable/unstable large and unstable small black holes with $C_P<0$, and an isolated branch of stable intermediate-sized black holes with $C_P > 0$. The reverse Van der Waals behaviour consists of a first-order phase transition is between a large stable black holes with $C_P > 0$ and a small unstable black holes with $C_P<0$; the intermediate-sized black holes also have $C_P<0$.

We also find that these black holes provide counterexamples to stability conjectures posited earlier~\cite{johnson2020,cong2019} that were based on the reverse isoperimetric inequality. It would be interesting to see if some generalization of these conjectures could be formulated (and proved) that would include the broader class of Ho\v{r}ava-Lifshitz black holes.

\appendix
\section{Smarr formula and the first law for the $k=0$ case}\label{app:k0case}

Here we present the thermodynamic relations of $z=4$ Ho\v{r}ava-Lifshitz black holes for the $k=0$ case. In this case the log term in the entropy \eqref{eqn:ent} vanishes and we find
\be
S=\frac{256 \pi ^2 P r_+^2}{3 \mu \kappa ^2 }. \label{eqn:app:ent}
\ee
We can substitute $r_+$ from this equation into eqs. \eqref{eqn:mass} and \eqref{eqn:temp}, along with $k=0$, to find the mass and temperature 
\begin{eqnarray}
	M &=& \frac{\sqrt{3}\kappa}{12\pi} \sqrt{\mu P}S^{3/2}, \label{eqn:app:mass}\\
	T &=& \frac{\sqrt{3}\kappa}{8\pi} \sqrt{\mu P}S^{1/2}, \label{eqn:app:temp}
\end{eqnarray}
which, like the entropy \eqref{eqn:app:ent}, are independent of the parameter $m$.

Putting $k=0$ in eqs. \eqref{eqn:volume} -- \eqref{eqn:y}, and using eq. \eqref{eqn:app:ent}, we find for the planar horizon case
\ba
V&=&\frac{\sqrt{3}\kappa}{24\pi} \sqrt{\frac{\mu}{P}}S^{3/2}, \qquad
H=0,\nn\\
U&=&\frac{\sqrt{3}\kappa}{24\pi} \sqrt{\frac{P}{\mu}}S^{3/2},\qquad
Y=\frac{\sqrt{3}}{12\pi} \sqrt{\mu P}S^{3/2}.
\ea
Therefore, the Smarr relation takes the form
\begin{equation}
	M=2ST - 2PV - \mu U - \frac{1}{2}\kappa Y,
\end{equation}
which can also be found by using the scaling relation of the mass \eqref{eqn:app:mass}. The first law  
\begin{equation}
	dM=TdS+VdP+Ud\mu+Yd\kappa ,
\end{equation}
is also easily shown to be satisfied.

\section*{Acknowledgements}

This work was supported in part by the Natural Sciences and Engineering Research Council of Canada.

\bibliography{mybib}
\end{document}